\documentclass[acmsmall,nonacm]{acm/acmart}

\usepackage{mylookup}
\usepackage{comment}
\usepackage{xspace}
\usepackage{dirtytalk}

\usepackage{enumitem}
\usepackage{array}
\setitemize{noitemsep,topsep=0pt,parsep=0pt,partopsep=0pt}
\usepackage{multirow}
\usepackage{xargs}

\AtBeginDocument{%
  }

\newcommand{\specialcell}[2][c]{\begin{tabular}[#1]{@{}l@{}}#2\end{tabular}}

\theoremstyle{definition}
\newtheorem{definition}{Definition}

\newtheoremstyle{property}
{1mm}                
{1mm}                
{\slshape}        
{}                
{\bfseries}       
{.}               
{ }               
{}                
\theoremstyle{property}
\newtheorem{property}{Block Proposal Property}

\newcommand{\positiveproof}{$[+]$\xspace}
\newcommand{\negativeproof}{$[-]$\xspace}
\newcommand{\positiveconj}{$[\pm]$\xspace}
\newcommand{\negativeconj}{$[\mp]$\xspace}
\newcommand{\notconj}{$[=]$\xspace}

\newcommandx{\btp}[5][1=none,4=dummy,5=dummy]{%
\IfStrEq{#4}{dummy}%
{\def\labeltext{\ref{bpp:#2}#3}}%
{\def\labeltext{\ref{bpp:#2}#3,\ref{bpp:#4}#5}}%
\IfEqCase{#1}{%
        {bold}{\textbf{BTP~\labeltext\xspace (\textit{\btpget{#2}{#3}})}}%
        {noname}{BTP~\labeltext}%
        {none}{BTP~\labeltext\xspace (\textit{\btpget{#2}{#3}})}%
    }[\PackageError{btp}{Undefined option to btp: #1}{}]%
}
\newcommand{\btpput}[3]{\lookupPut{\getrefnumber{bpp:#1}#2}{#3}}
\newcommand{\btpget}[2]{\lookupGet{\getrefnumber{bpp:#1}#2}}

\begin{document}

\title{On Replacing Cryptopuzzles with Useful Computation in Blockchain Proof-of-Work Protocols}

\author{Andrea Merlina}
\email{andremer@ifi.uio.no}
\author{Thiago Garrett}
\email{thiagoga@ifi.uio.no}
\author{Roman Vitenberg}
\email{romanvi@ifi.uio.no}
\affiliation{%
  \institution{University of Oslo}
  \city{Oslo}
  \country{Norway}}

\renewcommand{\shortauthors}{Merlina, Garrett, Vitenberg}

\begin{abstract}%
Proof-of-Work (PoW) blockchains have emerged as a robust and effective consensus mechanism in open environments, leading to widespread deployment with numerous cryptocurrency platforms and substantial investments. However, the commonly deployed PoW implementations are all based on solving cryptographic puzzles. Researchers have been pursuing the compelling idea of replacing cryptopuzzles with useful computing tasks for over a decade, in face of the substantial computational capacity of blockchain networks and the global pursuit of a more sustainable IT infrastructure. 
%
In this study, we conduct a comprehensive analysis of the prerequisites for alternative classes of tasks. We provide insight into the effect of introducing ``usefulness'' and of transitioning to task classes other than cryptopuzzles. Having distilled the prerequisites, we use them to examine proposed designs from existing literature. Finally, we discuss pertinent techniques and present research gaps in the current state-of-the-art.
\end{abstract}


\maketitle

\section{Introduction}

Since Bitcoin's~\cite{bitcoinpaper} inception in 2008, the system has attracted numerous stakeholders, growing in size and reaching a market cap of over 400 billion dollars~\cite{bitcoin-market-cap}. A central element in Bitcoin's design is the Proof-of-Work (PoW) mechanism for block proposals. PoW secures the chain, limits Denial-of-Service attacks, and improves the consensus algorithm scalability with respect to the participants. However, as it has been widely publicized and criticized, the energy consumption of PoW in Bitcoin is comparable to medium-sized countries. Indeed, at the time of writing, Bitcoin consumes as much annual energy as Belgium~\cite{cbeci}. While a number of alternative and less energy-consuming mechanisms have been proposed~\cite{proof-of-stake}, PoW remains a dominant mechanism in permissionless blockchains.

What makes the PoW mechanism in Bitcoin and other blockchain systems particularly wasteful is that it is based on solving cryptopuzzles. A solution to a cryptopuzzle does not serve any purpose or has any value outside of the blockchain ecosystem; one could argue that the hard-sought-after solutions to cryptopuzzle do not benefit humanity whatsoever. At the same time, in other contexts, a vast number of computational tasks need to be solved, whether in operational research, computational biology, Machine Learning, or other optimization problems.

The alluring dream of replacing useless cryptopuzzles with useful work has been tantalizing researchers for a while, motivating a large body of work in the area. However, despite numerous attempts and important academic contributions, no such system has been deployed yet. Where does state-of-the-art fall short, and can we still hope for such a replacement? Our paper aims to address this question systematically. 

Cryptopuzzles are exploited by the block proposal mechanism in Bitcoin and other systems where PoW grants the miner the right to propose a block. The block proposal mechanism plays a central role in ensuring correct system operation expressed through the well-known holistic requirements such as Common prefix, Chain quality, and Chain growth~\cite{garay}. We start by identifying the block proposal properties (BPPs) that lead to fulfilling the holistic requirements. Subsequently, we identify those properties of cryptopuzzles that are exploited by the block proposal mechanism toward satisfying the formulated BPPs. We observe that neither BPPs nor cryptopuzzle properties exploited in Bitcoin have been covered in the literature in a comprehensive or standardized way, yet each individual proposed scheme has been presenting a partial list using different specifications. To the best of our knowledge, this is the first work to present a comprehensive list of BPPs and exploited cryptopuzzle properties (including those that have never been mentioned in the literature) and analyze them in detail. In order to create a uniform framework, we isolate the PoW component in Bitcoin, present the component's interface, and use the interface to express the properties. For each property, we explain why it is needed for BPPs, how it is satisfied by cryptopuzzles, and how challenging it would be to satisfy it when using other classes of tasks. Although we do not prove our list of properties to be complete, it is more comprehensive than the previously known set of properties and can be used to assess the feasibility of alternative classes of tasks and novel PoW-based designs.

This property analysis allows us to consider the transition to a PoW system that is based on tasks other than cryptopuzzles. We discuss how the blockchain architecture and the interface of the PoW component need to be adapted to this end. We present two alternative definitions of ``usefulness'' in the context of PoW tasks and the need for a task supply model. Then, we systematically analyze the challenges posed by replacing cryptopuzzles with general classes of useful tasks. A number of those challenges are due to the introduction of usefulness, whereas the other challenges are encountered because some of the cryptopuzzle properties are difficult to replicate with other classes of tasks.

In light of this challenge analysis, we survey a range of proposals in the literature. To this end, we identify three popular classes of tasks (based on the problems of k-Orthogonal Vectors, Traveling Salesman, and Deep Learning, respectively), which have been touted as candidates for replacing cryptopuzzles in past proposals. After considering how well each class suits each PoW-induced requirement, we provide systematic coverage of individual proposals.

Our analysis reveals that usefulness is in direct conflict with many important properties. Moreover, no alternative task or system design that can satisfy all properties and completely replace cryptopuzzles has been proposed.

In summary, the contributions of this paper are as follows:
\begin{enumerate}
\item We systematically define, categorize, and discuss important properties that characterize blockchain tasks such as cryptopuzzles. The resulting framework of properties and the related discussion are of interest to both researchers and developers as a blueprint for the analysis and design of new systems. 
\item We consider the effect of introducing ``usefulness'' and of transitioning to task classes other than cryptopuzzles and analyze the implications for the blockchain architecture and for the identified properties.
\item We categorize classes of tasks considered in the literature as candidates for replacing cryptopuzzles and provide an in-depth analysis of the state-of-the-art for three of these classes.
\item We review and classify several proposed designs in the literature with respect to the identified properties, showing in practice how the proposed property framework can be used as a guide for both blockchain developers and researchers to assess the effectiveness of proposed designs. 
\item Finally, we identify research gaps in the state-of-the-art and outline a number of future research directions.
\end{enumerate}

The paper is structured as follows. In Section~\ref{sec:background}, we provide the necessary background on computational tasks, Bitcoin, and PoW. In Section~\ref{sec:properties}, we define and analyze the properties of classes of tasks that are necessary to support identified PoW requirements. After presenting the implications of usefulness in the context of PoW in Section~\ref{sec:problem}, we describe three popular classes of tasks in Section~\ref{sec:classes_of_tasks} and consider how well they satisfy the defined properties. This is followed by an analysis of several proposed designs in Section~\ref{sec:existing_approaches}. We present a list of research gaps in Section~\ref{sec:outlook}, cover related work in Section~\ref{sec:related_work}, and conclude in Section~\ref{sec:conclusion}.

\section{Background}
\label{sec:background}

Section~\ref{subsec:tasks} presents definitions for computational tasks. Next, Section~\ref{subsec:btc} presents a high-level overview of Bitcoin (the first and perhaps the most representative PoW-based blockchain system), which is used as a base model throughout this work. Section~\ref{subsec:btcpow} then describes PoW and cryptopuzzles in Bitcoin. Finally, we present the formal requirements of the consensus problem solved by Bitcoin in Section~\ref{subsec:btcproblem}.

\subsection{Computational Tasks}
\label{subsec:tasks}

A \textit{task} is a problem requiring non-trivial computation to produce a solution. 
In this work, such non-triviality is often referred to as \textit{hardness}. 
A \textit{class of tasks} is the type of problem being solved.
Examples of classes of tasks include linear algebra operations, satisfiability (SAT), and the Traveling Salesman Problem~\cite{cormen} (TSP).
Thus a task consists of a specific instance of a class of tasks, e.g. a specific weighted graph configuration in the case of TSP.
Solving a task implies executing a solution algorithm to find a solution to the task instance.
Algorithms for solving different classes of tasks have different computational complexity, as extensively surveyed in~\cite{intractability}.

In the context of this work, solving a task produces a Proof of Computation (PoC) that serves two purposes: (a) a solution to the task can be efficiently derived from the PoC, and (b) the PoC serves as a proof allowing an external party which did not take part in solving the task to verify that the task was solved correctly. We call the process in (b) task verification.

Next, we briefly provide a formal definition of three relevant classes of tasks.

\subsubsection*{k-Orthogonal Vectors}
Linear algebra operations, among which the calculation of orthogonal vectors, find applications in many disciplines, such as traffic flow and electric circuits. The k-Orthogonal Vector (k-OV) problem is defined as follows~\cite{ball1}.

\begin{definition}[k-Orthogonal Vectors]
The k-OV problem on vector of dimension $d$ is to determine, given $k$ sets ($U_1$,... $U_k$) of $n$ vectors in the form \{0,1\}$^{d(n)}$, which $u^s \in U_S$ for each $s \in [k]$ such that over $\mathbb{Z}$:

\begin{equation}
\label{eq:ov}
\sum_{l \in [d(n)]} u_l^1 \cdots u_l^k = 0    
\end{equation}

\end{definition}

The solution to a k-OV task is, therefore, a list of OVs.
No reduction is known from OV to CNF-SAT, and it is commonly conjectured that any algorithm requires $n^{k-o(1)}$~\cite{williams}. For the restricted case of $k = 2$, the complexity becomes $n^{2-o(1)}$.
Obtaining a truly sub-quadratic algorithm for OV has been elusive, and~\cite{williams2} showed that the Strong Exponential Time Hypothesis (SETH)~\cite{seth} implies the nonexistence of such an algorithm.

\subsubsection*{Traveling Salesman Problem}
Solutions to TSP tasks find applications in areas such as logistics by planning the shortest route among interest points, minimizing the cost of circuit board printing, as well as planning the movements of bulky astronomy telescopes, which are slow to re-target.
TSP is a classic example of an NP-complete problem for which, due to its hardness, there are several longstanding instances whose optimal solutions are not known~\cite{worldtsp, hard_tsp}.
It is defined as follows~\cite{gomes}.

\begin{definition}[Traveling Salesman Problem]

Given a set $C=\{c_1,...c_n\}$ of \textit{cities}, and a distance function $d : C \times C \rightarrow \mathbb{N}$, the solution to the Traveling Salesman Problem is a cycle that visits every city once, also known as a Hamiltonian cycle, resulting from a permutation of the cities $\pi : \{1,... ,n\} \rightarrow \{1,... ,n\}$, that minimizes the cost of traveling expressed as:

\begin{equation}
\label{eq:tsp}
\sum_{i=1}^{n-1} d(c_{\pi(i)}, c_{\pi(i+1)})
\end{equation}

\end{definition}

The Traveling Salesman Problem can be formulated as a decision problem by introducing a threshold $t_d$ so that a solution is valid if and only if the cost of the valid permutation $d_\pi$ is such that $d_\pi \leq t_d$.

The fastest exact solver of TSP is considered to be Concorde~\cite{concorde}. On the other hand, several heuristic algorithms such as LK~\cite{lk} and LKH~\cite{lkh} offer quicker and generally good quality solutions~\cite{lkh3}. 
A standard benchmark for testing those algorithms is the instance library TSPLIB~\cite{tsplib}. TSPLIB introduces a format~\cite{tsplibformat} as well, by which cities are identified in two-dimensional coordinates, and every distance pair is calculated with the Pythagorean theorem.

\subsubsection*{Deep Learning}
\label{subsec:deep_learning}
Machine Learning (ML) investigates methods that leverage data to improve the performance of a set of tasks. While the first ML methods date back to 1960, improved data access and hardware performance paved the way for more complex methods such as Deep Learning (DL), which is the subject of extensive research due to its successful track in tackling advanced tasks spanning from computer vision to machine translation. 

A Neural Network (NN) is a mathematical model of functional transformations often represented as a directed graph where edges have tunable weights and vertices, typically referred to as neurons, perform linear and non-linear transformations. NNs have been successfully employed in various applications, including supervised learning, which is described as follows.
Given a training set of $N$ example input-output pairs $(x_1,y_1), (x_2, y_2), ... (x_N, y_N)$, where each pair is generated by an unknown function $y=f(x)$, the task of supervised learning is to discover a function $h_w$ that approximate the true function $f$~\cite{aimodernapproach}.
Deep Learning (DL) encompasses a broad family of techniques in which NN layers are arranged in sufficiently significant numbers, hence the name Deep Neural Network (DNN). DL tasks are defined as follows~\cite{bishop}.

\begin{definition}[Deep Learning Task]
A Deep Learning Task consists of the process of learning a function $h_w:X\rightarrow Y$, where $h_w$ is parameterized by a vector $w \in W$ representing the tunable weights. The performance of $h_w$ on an example $(x_i, y_i)$ is measured by an error function $E(w)$ calculated across all input-output pairs. Therefore the goal is to minimize the following: 

\begin{equation}
\label{eq:ml}
E(w) = \frac{1}{2}\sum_{n=1}^{N}\parallel h_w(x_n, w) - f(x_n) \parallel ^{2}
\end{equation}

The error $E(w)$ is a smooth continuous function of $w$ that assumes the smallest value when the gradient vanishes, i.e. $\nabla E(w) = 0$.
Stochastic Gradient Descent (SGD) techniques make an update to the weight vector based on one data point at a time so that:

\begin{equation}
\label{eq:ml2}
w^{(\tau +1)} = w^{(\tau)} - \eta \nabla E(w^{(\tau)})
\end{equation}

where the parameter $\eta > 0$ is known as the \textit{learning rate}.
After each update, the gradient in Equation~\ref{eq:ml2} is re-evaluated for the new weight vector and the process repeats, in what is commonly referred to as epochs.
In the rest of the work we refer to NN error, performance and accuracy interchangeably, to express how close $h_w$ approximates the true function $f$.

\end{definition}

\subsection{Bitcoin}
\label{subsec:btc}

In Bitcoin, multiple nodes exchange messages over a partially synchronous and sparse peer-to-peer (P2P) network.
The system is \textit{permissionless} so that nodes are free to join or leave, and the complete set of participating nodes is unknown. Messages are broadcast following a gossip-based protocol~\cite{decker}. An asymmetric cryptography scheme is assumed by which all messages are signed to ensure authenticity and integrity. There are two types of messages: transactions and blocks. A transaction transfers the ownership of a resource from a sender to a recipient, both represented by public keys. A block is a data structure that contains an ordered set of valid transactions in addition to metadata. A valid transaction transfers resources that were not transferred before -- an attempt to transfer the same resource more than once is called \textit{double-spending} and is considered to be an attack on the system. The nodes creating transactions to be included in blocks are called \textit{clients}. The system's goal is to maintain an ordered, durable, and replicated history of blocks -- a blockchain, described next.

A blockchain is an ordered and immutable sequence of blocks, and it is a type of distributed ledger. The system maintaining a blockchain, such as Bitcoin, is called a \textit{blockchain system}. The position of each block in the blockchain is called \textit{block height}. Each block contains in its metadata the hash value of the previous block -- Bitcoin, for instance, uses the SHA-256 hash function. This cryptographic pointer to the previous block ensures that any modification in a block will modify the hash value of all following blocks in the chain. Thus in this work, we only consider a linear chain of blocks instead of other alternative structures such as directed acyclic graphs~\cite{dag_survey}. Blocks may be created by any node participating in the system. But for a block to be valid, its metadata must also include proof that the corresponding node had the right to create the block. Nodes compete for the right to create and append the next block of the blockchain by solving a cryptopuzzle, following a PoW consensus protocol -- the cryptopuzzle and how the consensus protocol handles conflicting block proposals are described later in this section. Nodes participating in the block proposal competition are called \textit{miners}. A miner that \textit{appends} a block (i.e. creates a block that eventually becomes part of the blockchain) is rewarded by financial means: the miner inserts a special self-rewarding transaction called \textit{coinbase transaction} in the proposed block. Nodes verifying the validity of blocks are called \textit{validators} -- all miners are also validators.

A certain fraction of nodes can behave arbitrarily, i.e. the presence of Byzantine faults is assumed~\cite{byzantine_generals}. The PoW protocol requires nodes to spend computational power to solve cryptopuzzles. In this context, an essential security assumption states that the honest nodes constitute more than half of the total computational power of the system. The immutability of the blockchain relies on this assumption. The specific cryptopuzzle that needs to be solved for each block depends on the hash value of the block, as described in Section~\ref{subsec:btcpow}. Thus to modify a block (e.g. to double-spend a resource), the PoW for the modified block and all following blocks must be recomputed faster than the rate at which the rest of the system creates new blocks. This is shown to be unfeasible under the security assumption above~\cite{bitcoinpaper}. To incentivize nodes to join the system and behave honestly (i.e. following the standard protocol), a miner receives two rewards for appending a block to the blockchain: block reward and transaction fees. Block reward consists of newly created resources given to a miner that created an appended block. At the same time, a transaction fee is a payment by the creator of a transaction to the miner that includes the transaction in an appended block. Moreover, transaction fees are a protection mechanism against Denial-of-Service (DoS) attacks~\cite{attacks_survey} since creating many transactions to flood the network and prevent nodes from performing tasks other than validating transactions would be too costly due to the fees.

\begin{figure}
    \centering

    \begin{minipage}{.5\textwidth}
      \centering
      \includegraphics[width=.7\linewidth]{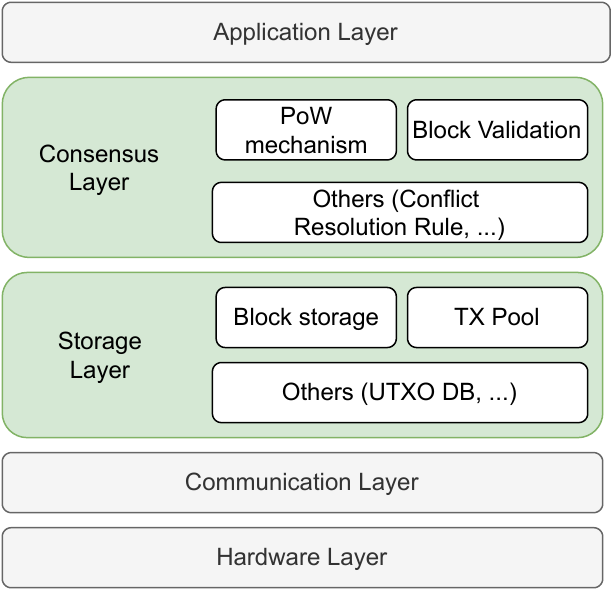}
      \captionof{figure}{Bitcoin layers.}
      \label{fig:blockchain_layer}
    \end{minipage}%
    \begin{minipage}{.5\textwidth}
      \centering
      \includegraphics[width=.9\linewidth]{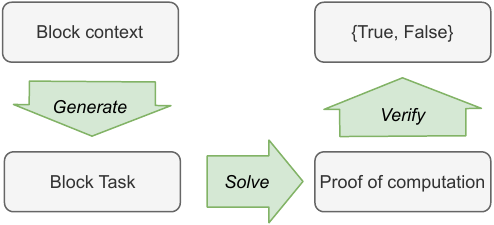}
      \captionof{figure}{Base interface. Boxes represent the inputs and outputs of the interface primitives, shown as arrows.}
      \label{fig:base_interface}
    \end{minipage}
    
\end{figure}

The Bitcoin model described in this section may be divided into five layers, as shown in Figure~\ref{fig:blockchain_layer}: Hardware, Communication, Storage, Consensus, and Application. The Hardware layer consists of the physical devices on which software implementing the blockchain system runs. In Bitcoin, it is common to employ ASICs and purpose-built hardware for solving cryptopuzzles, in addition to general-purpose hardware for other uses, such as storing the blockchain and communicating with other nodes. The Communication layer is responsible for connecting nodes in an overlay P2P network, on top of which nodes broadcast messages following a gossip-based broadcast protocol. The Storage layer deals with storing the blockchain and other data necessary for validating transactions and creating blocks. The Consensus layer guarantees that all honest nodes eventually agree on a single valid blockchain. Finally, the Application layer comprises the services that can be built using the blockchain system. Bitcoin, for example, supports payment applications, such as Lightning~\cite{lightning}. For more information on these layers, we refer the reader to~\cite{bitcoinbook,networking_survey,consensus_survey}. This work focuses on the Storage and Consensus layers, further described below.

In Bitcoin, many components are maintained by the Storage layer. In this work, we focus on two: Block storage and Transaction (TX) Pool. The \textit{Block storage} component stores valid blocks that were received or created by the node. This includes the blockchain and potentially conflicting blocks currently not part of the blockchain, as per the consensus protocol -- the \textit{orphaned blocks}. Note that a node may store only a subset of the blockchain. The \textit{TX Pool} stores all pending transactions, i.e. transactions not included in any proposed block yet. Each transaction contains a list of inputs and a list of outputs. Inputs are the outputs from other past transactions, which the current transaction is spending (i.e., transferring the corresponding resources). Outputs represent the recipients of the transferred resources (which come from the inputs). Examples of other components in the Storage layer include the Unspent TX Output Database (UTXO DB). This component stores unspent outputs i.e. outputs that were not used as inputs in any transaction already included in the blockchain. This component is essential for efficiently validating transactions: a transaction is only valid if its inputs are in the UTXO DB.

In this work, we focus on two components maintained by the Consensus layer: the PoW mechanism and Block Validation. The \textit{PoW mechanism} is responsible for producing a PoW (by spending computational resources) for a miner to be able to propose a valid block. Section~\ref{subsec:btcpow} describes this mechanism in more detail. A block proposed by a miner is broadcast to all other nodes, which validate the block through the \textit{Block Validation} component. For a block to be valid, it must contain: (i) the hash of the previous block in the chain; (ii) a set of valid transactions; and (iii) a valid PoW, i.e. the solution for the correct instance of a cryptopuzzle. Upon receiving a valid block, an honest node adds the block to its local blockchain replica. Since these two components are executed simultaneously by multiple nodes in the system, several blocks pointing to the same previous block may be proposed. This creates a \textit{fork} in the blockchain, resulting in multiple concurrent \textit{branches}. Different miners may continue proposing blocks extending different branches. Therefore, in the case of a fork, there are many different versions of the blockchain. Reaching consensus in this scenario means that all honest nodes agree on which branch is the main chain, discarding the blocks in the other branches (which become \textit{orphaned blocks}). This agreement is achieved through the Conflict Resolution component. In Bitcoin, this component enforces the \textit{longest chain} rule, which considers the branch with the largest number of blocks as the main chain. The rationale is that one of the concurrent branches will eventually become longer than the others due to the time variability of the PoW mechanism, described later in this work. In this scenario, increasingly more miners extend the same branch and \textit{converge} to the same chain. The longer such branch becomes, the lower the probability that another branch becomes longer. Therefore, the consensus in Bitcoin is probabilistic rather than deterministic~\cite{bft_vs_pow}. Although in this work we do not focus on the Conflict Resolution component, it is included, together with PoW mechanism and Block Validation, in the so-called Bitcoin \textit{backbone}, i.e. how blocks are proposed, validated, and accepted by nodes in the system to build the blockchain. The consensus problem solved by this backbone is further described next in Section~\ref{subsec:btcproblem}.

\subsection{PoW and Cryptopuzzles in Bitcoin}
\label{subsec:btcpow}

In a Proof-of-Work (PoW) protocol~\cite{pudding}, an entity (the prover) demonstrates to another entity (the verifier) that a certain amount of computational work was performed. PoW is a well-established mechanism to prevent Sybil and DoS attacks since PoW makes it computationally expensive for an attacker to create many identities and/or requests. In Bitcoin, PoW is also crucial for regulating the rate at which blocks are proposed by miners, in addition to DoS and Sybil protection. It takes, on average, 10 minutes for at least one miner in Bitcoin to propose a block containing a valid PoW. The importance of controlling the block proposal rate is further elaborated in Section~\ref{sec:properties}.

The PoW protocol in Bitcoin consists of solving a cryptopuzzle based on Hashcash~\cite{hashcash} and embedding the solution in the proposed block. Bitcoin's PoW finds a value -- the \textit{nonce} -- such that the result of a hash function applied to the concatenation of the nonce with a given input data is smaller than a pre-defined target value -- the \textit{difficulty}. Therefore, solving the cryptopuzzle for a given input data involves sequentially computing a hash function, varying only the nonce. The proof consists of the input data and the nonce, which can be verified efficiently by computing the hash function only once and comparing the resulting value with the expected difficulty. In the context of this work, Bitcoin's cryptopuzzle is a class of tasks. Historically, the use of hash functions to prevent content spamming was first proposed by Dwork and Naor in 1993~\cite{pow}.


The input data for a cryptopuzzle includes two values: (i) the hash of the root of the Merkle tree containing the set of transactions to be included in the block being created; and (ii) the hash of the previous block in the chain. We call this input data \textit{block context}. Merkle tree is the data structure used in Bitcoin to organize the transactions included in a block, which are taken from the TX Pool. To be able to propose a newly created block, a miner tries to solve the cryptopuzzle task. The proposed block thus contains the block context, the nonce, and the Merkle tree. The process of solving a cryptopuzzle to be able to propose a block is often called \textit{mining}.

Cryptopuzzles have some unique features when compared to other classes of tasks. Given the nature of cryptographic hash functions, the best available algorithm to solve cryptopuzzles is to try, in a brute-force fashion, many possible values for the nonce. The probability of any nonce value to be the solution is the same, and the solving process is \textit{memoryless}, i.e. the probability of any nonce value to be the solution does not change, regardless of how many other values have already been evaluated.
Typically, miners try so many nonces that Bernoulli trials, a discrete probability process, can be well approximated by a Poisson process, which is instead continuous. At the network level, the resulting time between block proposals is exponentially distributed, as formally shown in~\cite{bobtail}.
In Bitcoin, the nonce is a 32-bits value. In case a miner evaluates all $2^{32}$ possible values for the nonce without solving the cryptopuzzle, the miner may then modify the block context by, for example, changing the order of some transactions in the Merkle tree of the block being mined, resulting in a new cryptopuzzle instance to be solved. In that sense, a cryptopuzzle is considered computationally intensive yet solvable. We further discuss many properties satisfied by cryptopuzzles in the context of this work next in Section~\ref{sec:properties}.

\subsection{The Problem Solved by the Bitcoin Backbone}
\label{subsec:btcproblem}

In this work, we identify several properties related to cryptopuzzles in Bitcoin and map such properties to computational tasks in general. As a foundation for these properties, formal definitions of the requirements of the Bitcoin backbone are necessary. At a high level, Bitcoin solves the consensus problem in an open and permissionless environment. The formalization of this problem is not a trivial task by any means; it has been explored in several comprehensive and representative works~\cite{garay, pass, greeks, sz15}. At the level of the ledger abstraction, the two main properties are (a) persistence: if a transaction gets added to the public ledger, it never gets removed and (b) liveness: if an honest node wants to add a transaction to the ledger, the transaction should eventually be included in it~\cite{garay, pass}. However, the PoW protocol does not consider individual transactions or metadata, blocks are secured as atomic entities. Therefore, we do not consider the ledger abstraction in this work; instead, we focus on the underlying \textit{Bitcoin backbone abstraction} and its properties, which are more relevant to the problem of replacing cryptopuzzles with other tasks.

The Bitcoin backbone problem includes the following requirements~\cite{pass}:

\begin{description}

\item[Common Prefix:] at any point, the chains of two honest nodes can differ only in the last $T$ blocks with overwhelming probability, which depends on the parameter $T$;

\item[Future Self-Consistency:] at any two points $r$, $s$ during the execution, the chains of any honest node at $r$ and $s$ differ only within the last $T$ blocks with overwhelming probability, which depends on the parameter $T$;

\item[Chain Quality:] for any $T$ consecutive blocks in any chain held by some honest node, the fraction of blocks that were contributed by
honest miners is at least $\mu$ with overwhelming probability, which depends of the parameters $T$ and $\mu$;

\item[Chain Growth:] at any point in the execution, the chain of any honest node grows at a rate that does not fall below the ``minimum rate'' with overwhelming probability; the parameters of minimum rate and overwhelming probability need to be further specified, but this specification is not germane to the problem of replacing cryptopuzzles.
\end{description}

Intuitively, the properties of Common Prefix, Future Self-Consistency, and Chain Quality contribute to the safety and security of the Bitcoin system: they stipulate that bad scenarios never occur and that important invariants are never violated, even under attacks. On the other hand, chain growth is vital for a stable progress rate in Bitcoin.

The authors of~\cite{garay} and~\cite{pass} prove that the Bitcoin implementation solves the backbone problem under a number of specific assumptions about the environment in which the system operates and about the computational capabilities of the nodes. Since replacing cryptopuzzles with useful computation does not affect those assumptions and other aspects of the operational environment, discussing these assumptions is beyond the scope of our work.

The scope of formalization in the Bitcoin backbone problem is somewhat narrower than the entire Bitcoin system, as fairly observed by the authors of~\cite{garay}. In particular, the backbone problem does not consider node incentives due to the assumption that all nodes are statically divided into honest and malicious. In practice, the problem that Bitcoin solves has an additional requirement stipulating that the nodes are incentivized to participate in the system and that two computationally weaker miners do not gain higher rewards from combining their forces and acting as a single computationally stronger miner. This requirement prevents undesirable centralization in Bitcoin and boosts security in practice.

\section{Why cryptopuzzles work for PoW}
\label{sec:properties}

Despite the conceptual simplicity of PoW, cryptopuzzles and their use in the PoW mechanism satisfy surprisingly many properties. The goal of this section is to gain insight into these properties and show how they support the Bitcoin backbone protocol and other Bitcoin requirements listed in Section~\ref{subsec:btcproblem}.
Later, in Section~\ref{sec:problem}, we extend these properties to cover general computational tasks. We use the term \emph{Block Task} (BT) to refer to general computational tasks used to secure blocks; cryptopuzzles are, for instance, one example of BT.
Our methodology is as follows: first, we define a layer of properties of Block Proposals (BP) and show how they support Bitcoin requirements. Then, we define a layer of Block Task Properties (BTP) and show how they support Block Proposal Properties (BPP). 
In order to define the BTPs, we first formalize an interface for BTs in the context of PoW and associate the properties with individual primitives in the interface.

While BTPs and the Bitcoin requirements listed in Section~\ref{subsec:btcproblem} are presented formally, the properties of BP are stated at a more intuitive level. The goal of our paper is to motivate individual BTPs, connect them to Bitcoin requirements, and analyze how cryptopuzzles and other classes of tasks may satisfy these task properties. In the context of this work, BPPs mostly play a role in connecting task properties to Bitcoin requirements; it is not our goal to provide a specification for BPP.
To maintain a wide applicability, we employ generic terms such as ``too little'' or ``too much'' time. Quantification of these terms depends on the values of parameters in the backbone protocol, such as the Chain Growth rate. 
A precise definition would require specific assumptions on the system and models, which is outside the scope of this paper.
For the sake of readability, we group all properties into three categories. \emph{Safety and security} properties support the requirements of Common Prefix, Future Self-Consistency, and Chain Quality. \emph{Liveness} properties are used to guarantee Chain Growth while \emph{Decentralization} properties prevent the risk of centralization as described in Section~\ref{subsec:btcproblem}.
Table~\ref{table:table_properties_overview} gives an overview of the properties and their relation to the interface. 
Next, we start by defining the interface.


\begin{table}
\scriptsize 
\centering
\caption{Overview of Block Proposal and Block Task Properties}
\label{table:table_properties_overview}

\begin{tabular}{p{1.5cm}p{3cm}p{3cm}p{1.5cm}} 

\toprule
     \bf {\specialcell[t]{Classes of\\Properties}} & 
     \bf {\specialcell[t]{Block Proposal\\Properties (BPP)}} &
     \bf {\specialcell[t]{Block Task\\Properties (BTP)}} &
     \bf Interface\\
\toprule
    
& {\specialcell[]{Limited BP rate}} & {\specialcell[]{(a) BT hardness \\ (b) Context sensitivity \\ (c) Non-amortizability $^1$ \\ }}  & {\it \specialcell{Generate\\Generate\\Solve}}  \\ 

\cmidrule(lr){2-4}

& {\specialcell[]{Non-zero BP\\variability}} & {\specialcell[]{Non-zero variability\\across BTs}} & \textit{Solve} \\ 

\cmidrule(lr){2-4} 

{\specialcell[]{Safety \&\\ Security}} & {\specialcell[]{Adjustable lower \\ threshold for difficulty}} & {\specialcell[]{Adjustable \\lower threshold}} & {\it \specialcell[]{Generate}}  \\ 



\cmidrule(lr){2-4}

& {\specialcell[]{Block\\switchability}} & {\specialcell[]{(a) No reduction in solv.\\(b) Neg. BT gen. time\\(c) No incr. in sol. time}} & {\it \specialcell[]{Generate\\Generate\\Solve}}  \\ 

\cmidrule(lr){2-4}

& {\specialcell[]{BP verification soundness}} & {\specialcell[]{BT verification  soundness}} & {\it \specialcell[]{Verify}}  \\ 

\midrule

& {\specialcell[]{BP verification\\efficiency}} & {\specialcell[]{(a) BT generation efficiency $^2$\\ (b) BT verification efficiency}} & {\it \specialcell[]{Generate\\Verify}}  \\
\cmidrule(lr){2-4}
Liveness & {\specialcell[t]{BP timeliness}} & {\specialcell[]{(a) BT generation efficiency $^2$\\(b) BT solvability\\(c) BT tractability}} & {\it \specialcell[]{Generate\\Generate\\Solve}} \\
\cmidrule(lr){2-4}
& {\specialcell[]{Adjustable upper \\ threshold for difficulty}} & {\specialcell[]{Adjustable\\upper threshold}} & {\it \specialcell[]{Generate}}  \\ 
\cmidrule(lr){2-4}
& {\specialcell[]{BP verification completeness}} & {\specialcell[]{BT verification completeness}} & {\it \specialcell[]{Verify}}  \\

\midrule

Decentralization & BP fairness & {\specialcell[]{(a) No sup. dep. on res. \\ (b) Non-amortizability $^1$}} & {\it \specialcell[]{Solve\\Solve}} \\

\toprule

\end{tabular}

\tiny{$^1$ Same BTP for {\it Limited BP rate} and {\it Proposal fairness} }\\
\tiny{$^2$ Same BTP for {\it BP verification
efficiency} and {\it BP timeliness} }\\

\end{table}

\subsection{Block Task interface}
\label{sec:base_interface}

The life cycle of BTs in Bitcoin-based blockchains is as follows. A miner partially forms a new block, thereby creating a block context (see Section~\ref{subsec:btcpow}). Using the block context, the miner \textit{generates} an associated BT and attempts to \textit{solve} such BT. If successful, the miner inserts the solution into the previously formed block, creates a BP, and broadcasts the BP to other miners. Upon receiving the BP from the network, the validators also \textit{generate} a BT using the block context and \textit{verify} the validity of the solution included in the BP w.r.t. the BT.

All operations in the cycle are local, apart from broadcasting and receiving the BP from the network.
This cycle can be interrupted at any moment, e.g., by the arrival of new blocks from the network, which may result in updating the block context. In such scenarios, the cycle is started anew.
In the following, we focus on the generation, solution, and verification of BTs, leaving aside elements relevant to BPs but not for BTs, such as the selection of transactions to be included in the block or their validation.
Table~\ref{table:block_task_interface} and Figure~\ref{fig:base_interface} show the inputs and outputs of the three operations in the interface.

\begin{table}[t]
\newcolumntype{x}[1]{>{\centering\arraybackslash\hspace{0pt}}m{#1}}
\scriptsize
\centering
\caption{Block Task Interface}
\label{table:block_task_interface}
\begin{tabular}{cx{2in}x{1in}x{1.5in}}

& \textbf{Generate} & \textbf{Solve} & \textbf{Verify} \\ 
\toprule
\textbf{Input} & Block Context & Block Task & Block Task, Proof of Computation \\  \midrule
\textbf{Output} & Block Task & Proof of Computation & Boolean  \\ 
\midrule
\textbf{Descr.} & 
    Generates a Block Task such as a cryptopuzzle instance based on the block context. Generation is performed independently by every miner and validator. & 
    Search for a solution that satisfies the validity requirements. & 
    Verifies that the proof of computation (previously produced by Solve) is indeed a valid solution for the Block Task. \\

\toprule

\end{tabular}
\end{table}

\btpput{rate}{a}{BT hardness} 
\btpput{rate}{b}{Context sensitivity} 
\btpput{rate}{c}{Non-amortizability} 
\btpput{variability}{}{Non-zero variability across BTs} 
\btpput{adj}{}{Adjustable lower threshold} 
\btpput{switch}{a}{No reduction in solvability} 
\btpput{switch}{b}{Negligible BT generation time} 
\btpput{switch}{c}{No increase in solution time} 
\btpput{sound}{}{BT verification soundness} 
\btpput{verif}{a}{BT generation efficiency} 
\btpput{verif}{b}{BT verification efficiency} 
\btpput{timeliness}{a}{BT generation efficiency} 
\btpput{timeliness}{b}{BT solvability} 
\btpput{timeliness}{c}{BT tractability} 
\btpput{adj2}{}{Adjustable upper threshold} 
\btpput{compl}{}{BT verification completeness} 
\btpput{fairness}{a}{No superlinear dependency on the resources} 
\btpput{fairness}{b}{Non-amortizability} 

\subsection{Safety and Security Properties}
\label{sec:security_properties}

The high-level goal of safety and security properties is to prevent incorrect system states. 
In the context of blockchain, a system is secure if blocks are correctly appended to the chain (refer to the Common Prefix, Future Self-Consistency and Chain Quality properties). 
One particularly harmful scenario for safety and security is the proliferation of chain forks and increased branching (which in the following are used interchangeably) because forks and branching invalidate the Bitcoin backbone properties described in Section~\ref{subsec:btcproblem}.
With many forks, the probability of miners receiving blocks in different order increases. Therefore, the divergence of the chains of two honest miners probabilistically increases and the probability to violate Common Prefix by exceeding the last $T$ blocks is higher.
Similarly, high branching increases the risk of violating Future Self-Consistency.
Another effect of forks is reduced mining utilization~\cite{bitcoinng} i.e. reduced amount of computing power securing the chain. Lower mining utilization creates a risk of scenarios in which several consecutive blocks are appended by malicious nodes, invalidating the Chain Quality property.
Hence, increased branching harms the security and safety of blockchain.

In the following, we intuitively describe five important BPPs for security, namely
\textit{Limited Block Proposal rate}, \textit{Non-zero Block Proposal variability}, \textit{Adjustable lower threshold
for difficulty}, \textit{Block switchability} and \textit{Block Proposal verification soundness}.

\begin{property} [Limited Block Proposal Rate]
\label{bpp:rate}
The probability that a valid block proposal takes ``too little'' time to form is ``sufficiently'' small.
\end{property}

A fundamental requirement for safety is that blocks cannot be proposed or appended to the local copies of the ledger at too high of a rate. 
With a rate too high, the probability of branching increases, thereby affecting security.
In general, the proposal rate that maintains branching at an acceptable level depends on other parameters such as block size and information propagation mechanisms, as extensively studied in other works~\cite{gervais}. 
Besides, PoW-based blockchains may tolerate a small number of such fast proposals, but the probability of their creation needs to be low. 
We identified three BTPs relevant for BPP~\ref{bpp:rate}. 

\btp[bold]{rate}{a}: Given an arbitrary implementation $I$ of the \emph{Solve} function, the probability that \textit{Generate} returns a ``simple'' BT, that is a BT on which an invocation of $I$ takes too little time to run is low. It is obvious that the existence of many simple BTs invalidates BPP~\ref{bpp:rate}. 

\btp[bold]{rate}{b}: The BT produced by \emph{Generate} depends on the input block context. Specifically, given two different block contexts, the probability to \emph{Generate} the same BT for each is very low.
The rationale for \btp[noname]{rate}{b} is to prevent unwarranted proposals that are, for example, reusing already solved instances or stealing the solution of another miner.
Since it is possible to develop a dictionary of past BT solutions, \btp[noname]{rate}{b} strictly limits the possibility that miners submit valid proposals without performing any work. This would increase the rate of valid proposals, invalidating BPP~\ref{bpp:rate} regardless of the hardness of the reused BT.
In particular, this property means that the space of BTs produced by \emph{Generate} is ample.
Interestingly, \btp[noname]{rate}{b} does not protect from the case in which the miner can gain knowledge about the solution from outside the system. For example, there could be an external database of all known instances and solutions. Thus, the space of BTs produced by \emph{Generate} needs to be much larger than the space of known solved instances.
In Bitcoin, \btp[noname]{rate}{b} holds due to the enormous space of values that the block context might assume: the probability of generating the same BT is thus overwhelmingly low.

\btp[bold]{rate}{c}: \textit{Solve} cannot be amortized across a set of different BTs so that the process of solving BT~{\it A} provides no speed-up for the subsequent process of solving BT~{\it B}. 
If the solution process can be amortized, the time to create a valid BP may be reduced in practice, especially for a miner that has been solving BTs for a long time.
For example, computing a given Fibonacci number can be efficiently sped up by the knowledge of lower-value Fibonacci numbers.
In another example, the solution to the Orthogonal Vector problem is the number of vectors in a given set whose dot product is equal to zero.
Suppose multiple tasks include input sets with equal vectors. In that case, the computation across tasks can be amortized: knowing the solution for two sets of vectors removes the need to compute it again for the other sets.
Similarly, if some specific configurations are known to repeat, miners may build dictionaries of (intermediate) solutions to amortize the block proposal process. For instance, in Machine Learning, it is common practice to reuse semi-trained models to shorten the training time. 
The amortization issue applies to many types of tasks, and it is exacerbated in systems where external clients supply arbitrarily tasks.
In Bitcoin, on the other hand, \btp[noname]{rate}{c} is guaranteed by the nature of cryptopuzzles, for which the solution search is memoryless, and therefore no private state can speed up the search.


\begin{property} [Non-zero Block Proposal variability]
\label{bpp:variability}
Given an arbitrary chain, the times at which different miners produce competing BPs for the next block exhibit non-zero variability.
\end{property}

\begin{figure}
    \centering
    \includegraphics[width=.5\linewidth]{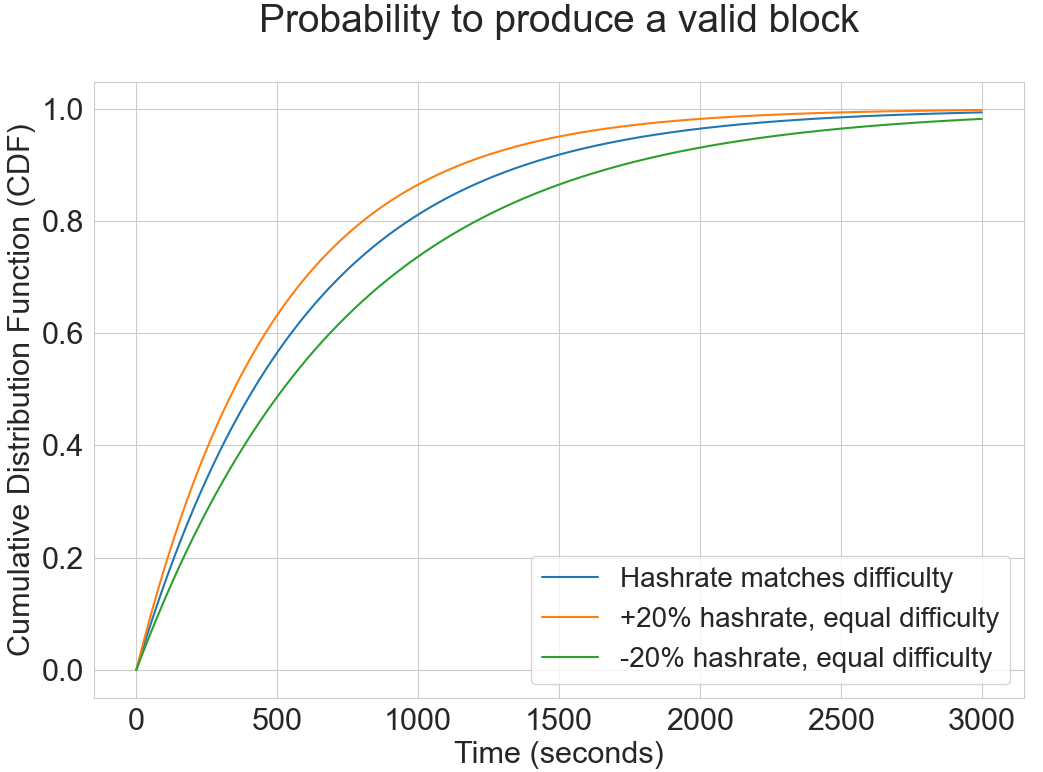}
    \caption{Block proposal probability in Bitcoin}
    \label{fig:bitcoin_probability}
\end{figure}

In PoW-based blockchains, BP formation is a stochastic process. Bitcoin's block proposal, for example, is modeled as a Poisson process~\cite{decker} whose time between blocks follows an exponential distribution with a 10 minutes mean value, as shown in Figure~\ref{fig:bitcoin_probability}. The variability of BP limits the number of conflicting blocks in the network at any time, which reduces the number of forks. 
Ideally, no two proposal solutions should be found at approximately the same time.
However, the amount of proposal variability needed for the desired level of security depends on several system parameters, such as the number of miners and propagation delay. 
One method to satisfy BPP~\ref{bpp:variability} is through sufficiently variable BT solution times, thus justifying the following property.

\btp[bold]{variability}{}: Let us call an implementation $I$ of the \emph{Solve} function optimal if no other implementation asymptotically outperforms $I$. 
Given an arbitrary optimal $I$, the number of instructions $I$ requires to solve a BT exhibits non-zero variability across different BTs of the same difficulty.

In Bitcoin, this property is satisfied due to the brute-force search of the cryptopuzzle solutions. Indeed, the distribution of times in Figure~\ref{fig:bitcoin_probability} shows significant variability. On the other hand, optimal algorithms to solve some classes of tasks terminate after a fixed number of operations. 
A simple example is matrix multiplication. An optimal algorithm for multiplying two matrices requires the same amount of computing steps regardless of the input and therefore does not satisfy \btp[noname]{variability}{}{}.
\btp[noname]{variability}{} is sufficient for fulfilling BPP~\ref{bpp:variability}, but not necessary. In general, BPP~\ref{bpp:variability} may be satisfied by means other than variable task solution times. For example, it can be achieved by slightly varying the difficulty of BTs assigned to different block contexts. Additionally, in real systems, several factors such as network delay and heterogeneous computing power increase the variability in which nodes receive block proposals. 

    
\begin{property} [Adjustable lower threshold for difficulty]
\label{bpp:adj}
When scaling up, the system efficiently and effectively controls a lower limit on the average time to propose valid blocks.
\end{property}

In permissionless blockchains, new nodes may join the systems or, alternatively, existing nodes may get upgraded thus increasing the aggregate computing power of the network.
Nonetheless, BPP~\ref{bpp:rate} must be preserved when the computing power varies dynamically over time. 
Therefore, the system should be able to dynamically adjust the difficulty of creating new blocks.

BPP~\ref{bpp:adj} is supported by the following \btp[bold]{adj}{}: it is possible to efficiently and effectively assess and increase the difficulty of BTs produced by \textit{Generate}.

In Bitcoin, for example, the BP rate is controlled by the cryptopuzzle's difficulty which is adjusted periodically every two weeks as follows.
The network profiles the time to create blocks over the two-week period. 
If more than 2016 blocks are proposed in that period, the difficulty is raised; if fewer, the difficulty is lowered~\cite{adj_algo}. 
Thus the system adjusts the difficulty such that a block is proposed on average every 10 minutes, given the estimated total mining power. 
 
Note that BPP~\ref{bpp:adj} focuses on scaling up. We explicitly distinguish between adjustability for scaling up and scaling down, as the latter affects liveness rather than security.


\begin{property}[Block Switchability]
\label{bpp:switch}
A miner receives, on average, a higher reward by accepting a valid incoming block that would alter the local blockchain copy according to the protocol, 
and mining on top of it rather than by aiming to propose a conflicting block.
\end{property}

In the Bitcoin core\footnote{https://github.com/bitcoin/bitcoin} protocol implementation, valid incoming blocks are appended to the local ledger and referenced in the following block. However, every miner is free to choose a different implementation so that miners should be incentivized to accept a valid incoming block and mine on top of it. 
Consider the case in which BTs are invariably solved after a fixed number of operations. 
Then, a miner close to producing a block may prefer to ignore incoming proposals and aim to produce conflicting blocks with the hope of winning the ensuing forks.
Such a strategy increases branching, with the security consequences described in Section~\ref{sec:security_properties}. 

In essence, BPP~\ref{bpp:switch} depends on the probability of existence of a valid block, on the average time for the miner to find a valid block, and on the probability of a proposed valid block to be included in the chain. 
There is a complex interplay between these three factors that may affect the miner's decision whether to switch. 
In general, if the difference in one factor is vast while the difference in the other two factors is slight, the differentiating factor will dominate the others.
For example, in cases when the probability of existence of a valid block in the new context is zero, the miner may prefer to reject the incoming block and continue mining on top of the old one instead. 

Ideally, the new block will be at least as good as the old block with respect to all three factors. This can be phrased as the following sufficient (but not necessary) conditions of BPP~\ref{bpp:switch}: (a) the probability of existence of a valid block with the new context is not reduced, (b) the expected time to propose a new block does not increase, and (c) the probability of a proposed valid block to be included into the chain does not decrease.
Condition (c) is external to BTs. On the other hand, conditions (a) and (b) are supported by the following BTPs. 

\btp[bold]{switch}{a}: when switching to the new BT produced by \textit{Generate} out of the new block context, the probability of existence of a valid solution for the new BT is not reduced compared to the BT produced by \textit{Generate} out of the old block context.
\btp[noname]{switch}{a} is necessary for condition (a) of BPP~\ref{bpp:switch} since the probability of existence of a valid proposal relies on the probability of existence of the solution to the BT.
\btp[noname]{switch}{a} is important for classes of tasks that may not have any valid solution and for which different BTs have a different probability of having a solution. For example, Bitcoin cryptopuzzles may not have any valid solution, but the probability for a cryptopuzzle to have a solution solely depends on its difficulty. Therefore, different cryptopuzzles of the same difficulty will have the same probability, thereby satisfying \btp[noname]{switch}{a}. 

\btp[bold]{switch}{b}: the time to \textit{Generate} a new BT is negligible compared to the time to \textit{Solve} a BT.
\btp[noname]{switch}{b} is related to condition (b) of BPP~\ref{bpp:switch}: 
a \textit{Generate} of non-negligible duration with respect to the time required to solve a BT, directly increases the expected time to propose a new block.

\btp[bold]{switch}{c}: the expected time for the \textit{Solve} function to find a valid solution for the new BT does not increase.
This property compares the expected time to solve a new BT with the expected time to solve the old BT. 
\btp[noname]{switch}{c} directly affects condition (b) because an increased time to \textit{Solve} the BT extends the expected time to propose a new block.

As mentioned above, a deterministic computation such as matrix multiplication is unfit in this regard because it is faster to complete an ongoing task rather than to start anew. \btp[noname]{switch}{c} is also important for classes of tasks that vary in terms of the expected time to find a solution. The property holds in Bitcoin because solving cryptopuzzles is a memoryless process. Hence the already performed computation does not affect the probability nor the expected time to find a solution. There is, however, one case where \btp[noname]{switch}{c} is violated in Bitcoin, namely when there is a difficulty increase and the next block has a higher difficulty. In this case, the miners may prefer not to switch and rather continue mining on top of the previous block. Interestingly, there have been no reports of an increased rate of forks in Bitcoin at points where the difficulty increases.

Besides blocks, miners receive transactions from the network. 
Similarly to the arrival of a new block, the arrival of a new transaction being included in the block changes the block context and the BT. 
Intuitively, it should not be detrimental for miners to include new transactions in the block while mining; otherwise, miners may not be incentivized to modify the BT they are currently solving, leading to longer transaction confirmation delays and smaller total fees rewarded to miners. 
In Bitcoin, for example, it is a common practice for miners to replace transactions with lower fees as described in BIP125~\cite{bip125}. However, in this work, we do not explicitly consider switchability due to incoming transactions as a property. Our focus is on the correctness requirements for blockchain, as defined in Section~\ref{subsec:btcproblem}, as opposed to performance optimizations and metrics.

    
\begin{property} [Block Proposal verification soundness]
\label{bpp:sound}
The BP verification protocol has a high probability of rejecting an invalid BP i.e. the verification is sound.
\end{property}
Every block proposal received from the network is first verified and then appended to the local copy of the ledger. In principle, the verification need not be deterministic. There exist several probabilistic verification protocols, such as zero-knowledge proofs~\cite{zkp}. Such types of verification have a probability of misclassifying invalid proposals as valid. If the probability of such false negatives is too high, many invalid blocks are included in the chain, negatively affecting the Chain Quality. 
Therefore, the probability of correct classification must be high, as stated in the following property.

\btp[bold]{sound}{}{}: \textit{Verify} has a high probability of rejecting an invalid BT solution.

In Bitcoin, 
the verification is deterministic: it suffices to verify that the block hash is lower than the target value. Therefore it is certain that a valid solution is classified as such.


\subsection{Liveness Properties}
\label{sec:liveness_properties}

Liveness in distributed systems has been extensively studied in several important works~\cite{lynch}. In the context of blockchain, liveness has been expressed in terms of Chain Growth and the rate at which blocks are appended to the ledger, as described in Section~\ref{subsec:btcproblem}. 
In the following, we list a number of properties related to liveness: \textit{Block Proposal verification efficiency}, \textit{Block Proposal timeliness}, \textit{Adjustable upper threshold for difficulty}, and \textit{Block Proposal verification completeness}.


\begin{property} [BP verification efficiency]
\label{bpp:verif}
Given a BP, miners can efficiently verify its validity. 
\end{property}

Inefficient verification impacts the system in several negative ways.
To begin with, inefficient verification slows down block propagation because miners validate a proposal before advertising it to the neighbors, as described in Section~\ref{sec:background}. Slower block propagation, in turn, delays the chain growth of honest nodes, possibly below the minimum allowed rate. In that sense, inefficient verification affects safety as well because slower block propagation increases the likelihood of branching. Besides, the need to invest non-negligible resources in verification may disincentivize the nodes from performing it.
BPP~\ref{bpp:verif} is supported by two BTPs.

\btp[bold]{verif}{a}: \textit{Generate} is efficient in creating a new BT. An efficient implementation of the generation primitive is central to block validation (as well as to other mechanisms) since \textit{Generate} is executed as part of the preliminary computation to verify that the BT solved in the given BP is correctly derived from the block context of that BP. In reality, there are several factors affecting the efficiency of task generation. For example, if the BT generation requires access to task metadata stored remotely, the generation time is directly affected by the fetching delay.

\btp[noname]{verif}{a} is similar to \btp[noname]{switch}{b}: both require BT generation to be efficient. However, \btp[noname]{verif}{a} is phrased as absolute efficiency in terms of the low runtime whereas \btp[noname]{switch}{b} stipulates relative efficiency compared to the solution time. In reality, it implies that \btp[noname]{verif}{a} is stronger than \btp[noname]{switch}{b} because the solution time is supposed to be significant according to \btp[noname]{rate}{a}. We nevertheless opt to introduce \btp[noname]{switch}{b} because there exist systems that satisfy \btp[noname]{switch}{b} but not necessarily \btp[noname]{verif}{a}, as we show in Section~\ref{sec:existing_approaches}.

\btp[bold]{verif}{b}: \textit{Verify} is efficient. Clearly, an inefficient BT validation makes the BP validation inefficient as well. Thus, both \btp[noname]{verif}{a} and \btp[noname]{verif}{b} are necessary for BPP~\ref{bpp:verif}.

The exact meaning of efficiency depends on the specific blockchain system and the class of tasks being solved. It typically implies both low asymptotic computational difficulty and low computation times in practice since the verification is performed multiple times by a large number of nodes, including those with lower computing power.

In Bitcoin, the implementation of both \emph{Generate} and \emph{Verify} only consists of a small number of instructions. In particular, \emph{Verify} entails a single hash computation and a comparison.

\begin{property} [Block Proposal timeliness]
\label{bpp:timeliness}
Given an arbitrary chain, the average time to produce a valid BP for the next block is ``sufficiently fast''.
Additionally, there is a very high probability of producing at least one valid BP for the next block within a time that is not ``exorbitantly slow''.
Both times are set so as to satisfy Chain Growth.
\end{property}

The performance of blockchain systems is characterized by several important metrics related to liveness, namely the number of transactions appended per second and the time it takes for a transaction to become part of the main chain with high probability. Since transactions are batched and appended in blocks, these metrics are affected by the timeliness of BP.
Without timely proposals, the performance of the system would be degraded, and the rate at which the chain grows may drop below the minimum rate required by the Chain Growth property. 

It is actually non-trivial to capture the exact requirement of BP timeliness that is necessary for the blockchain system in practice. Intuitively, it is the opposite of hardness expressed as BPP~\ref{bpp:rate}: while all concurrent block proposals should be sufficiently hard, at least one concurrent block should be proposed within a reasonable time. However, probabilistic artifacts are not as detrimental to hardness as to timeliness. Suppose once in a rare while, there is a block that is secured by a relatively small amount of work. In that case, it will not be a significant problem for the security of the entire chain because later blocks will additionally secure the block. On the other hand, a block taking weeks to produce would cause a significant problem for Chain Growth and liveness in general. For this reason, BPP~\ref{bpp:timeliness} specifies two probabilistic thresholds: one threshold to guarantee the average speed of appending blocks to the chain and another threshold to curb probabilistic artifacts.
BPP~\ref{bpp:timeliness} is supported by three BTPs. 
The first, \btp[bold]{timeliness}{a}, is equivalent to \btp[noname]{verif}{a}: \textit{Generate} must produce new BTs efficiently.
Since \textit{Generate} is invoked every time a miner creates a BT and verifies a BP, an efficient implementation is important for the timeliness of BPs. 

However, BTs may have no solution at all. For example, not all cryptopuzzles have a solution. A block cannot be secured by proof that the generated BT has no solution because neither cryptopuzzles nor most other task classes allow for such negative proof. This complication is addressed as follows in the Bitcoin design: \emph{Solve} returns a special marker signifying that no solution has been found. In this case, the miner changes the block context, e.g., by modifying the timestamp or rearranging the transactions. Such a change results in a new block context and, consequently, a different BT for the miner to solve.
This works, however, only if the following property is satisfied.

\btp[bold]{timeliness}{b}: a BT produced by \emph{Generate} has a sufficiently high probability of having a solution. The probabilistic threshold may depend on the adjustable difficulty of the task (see BPP~\ref{bpp:adj2}).
In Bitcoin headers, for instance, the difficulty is represented as a 4-byte value, thus spanning between 0 and $4.3 \cdot 10^9$; even with a difficulty of 1, there is a 36.7\% chance of not finding any valid hash in the entire $2^{32}$ nonce range. At the difficulty of one million, there is a 99.999905\% that there is no solution in the entire nonce range~\cite{probNoSol}.

Finally, BTs must allow for a sufficiently efficient implementation of \emph{Solve}, which either finds a solution or concludes that no solution exists. It is conceivable, however, for a miner to use an implementation of Solve that may return without finding a solution even when a solution does exist. Imagine a class of BTs that consists of two categories: BTs that can be solved by a more efficient method and BTs that cannot. A miner may opt to use an implementation of Solve that only tries the more efficient method because it may be faster to apply the more efficient method to multiple generated BTs than to apply the more expensive method to a single BT.
An efficient \emph{Solve} implementation limits both the average runtime and the risk of probabilistic artifacts, as discussed above.
A simple way to formulate this requirement is to consider the entire space of BTs producible by \emph{Generate}, regardless of the block context: 

\btp[bold]{timeliness}{c}: consider the entire set $S$ of BTs that \emph{Generate} may produce. Then, there exists an implementation $A$ of \emph{Solve} such that (a) $A$ finds a solution for a ``significant'' fraction of BTs in $S$, (b) the average runtime of $A$ over the BTs in $S$ is ``sufficiently fast'' and (c) the fraction of BTs in $S$ that takes $A$ ``exorbitantly high'' time to process is very low.
For cryptopuzzles, the best known \emph{Solve} implementation is a brute-force search over the nonce space. This implementation always finds a solution if one exists. Its average runtime depends on the size of the nonce space (which is fixed in Bitcoin and equal to 32 bits) and on the probability of each point in the nonce space being a solution, which depends on the adjustable difficulty. The worst runtime for a single cryptopuzzle is simply the time required to scan the entire nonce space and calculate the SHA-256 hash function for each point in the space. However, the process of creating a block proposal may require solving multiple BTs until a BT with a solution is generated, as described above. 

In Bitcoin, blocks are proposed every 10 minutes on average, whilst 99\% of the blocks are found within 2763 seconds i.e. in less than five times the average. So far, the longest block proposal time has been for the first proposed block, i.e., after the genesis block. It took more than five days. At that time, however, Nakamoto was most likely the only miner in the system. 
The second most prolonged delay was one day, one hour, and eight minutes. It took place between blocks 15323 and 15324~\cite{delayPost}.
On the other hand, if Bitcoin continues to operate for an extended time, a probabilistic artifact with a longer block creation time is likely to occur.

\btp[noname]{timeliness}{c}{} considers the entire set $S$ of BTs that \emph{Generate} may produce. In principle, a PoW system may consider the performance of \emph{Solve} for a more restricted set of BTs. For example, given an arbitrary chain and a set of block contexts for the next block, the system may consider the set $S$ of BTs that \emph{Generate} produces for each block context. In this case, it might be possible to consider only the best runtime of a \emph{Solve} implementation among all the tasks in this limited $S$. 
However, we are unaware of any research in the literature or any practical attempts to create such a PoW system.


\begin{property} [Adjustable upper threshold for difficulty]
\label{bpp:adj2}
When scaling down, the system efficiently and effectively controls an upper limit on the average time to propose valid blocks. 
\end{property}

This property is the counterpart of BPP~\ref{bpp:adj}.
When the number of blockchain participants decreases, the system should be able to adjust the difficulty to maintain the desired BP rate.
Without this property, the time required to produce BP would increase, impacting BPP~\ref{bpp:timeliness}. 
At the task level, we define the following property. 
\btp[bold]{adj2}{}: it is possible to efficiently and effectively assess and reduce the difficulty of BTs produced by \textit{Generate}.


\begin{property} [BP verification completeness]
\label{bpp:compl}
The BP verification has a low probability of rejecting a valid BP i.e. the verification is complete.
\end{property}

Considerations analogous to those presented for BPP~\ref{bpp:sound} apply to BPP~\ref{bpp:compl} as well.
In the case of probabilistic verification, it should be highly unlikely to reject a valid solution; otherwise, miners may discard valid proposals, negatively affecting the BP rate and, therefore, Chain Growth. The BT property is stated as follows.

\btp[bold]{compl}{}: The probability of \textit{Verify} rejecting a valid PoC is low.

In Bitcoin, the validation algorithm is deterministic, and therefore completeness is guaranteed.


\subsection{Decentralization Property}
\label{sec:decentralizations_properties}

It is critical that different blockchain nodes contribute blocks to the main chain in order to avoid centralization. This is achieved by incentivizing individual nodes to propose blocks through a system of rewards. However, rewards only work if each node has a fair chance of proposing a block that will be included in the main chain.
In this section, we discuss the fairness of rewards due to block creation and how the process of solving tasks directly impacts these rewards. 

\begin{property} [BP fairness]
\label{bpp:fairness}
The probability of proposing a block is proportional to the relative amount of resources spent by the node on generating the BP.
\end{property}

Assume that there are two miners, $A$ and $B$, with a computing power of $C_A$ and $C_B$, respectively. Further, assume that if $A$ and $B$ mine independently, their computing power grants them the respective probabilities of $P_A$ and $P_B$ to propose a block. Then, in order to disincentivize $A$ and $B$ from pooling their computing resources, the probability of $A$ and $B$ proposing a block if working together $P_{AB}$ needs to be $P_{AB} \leq P_A + P_B$. In other words, the probability of proposing a block relative to the computational power must follow a non-superlinear function.
Superlinearity is detrimental to decentralization because miners gain more than their proportional reward by joining forces or increasing their computational power. 

In reality, systems supporting BPP~\ref{bpp:fairness} (such as Bitcoin) exhibit a linear increase in the probability. It is an open question whether a sublinear increase is desirable, as it implies that every miner is incentivized to split itself into as many miners as possible. At the BT level, BPP~\ref{bpp:fairness} is formulated as follows.

\btp[bold]{fairness}{a}: The probability that \textit{Solve} produces a valid solution is proportional to the relative amount of resources spent by the node on executing \textit{Solve}.

\btp[noname]{fairness}{a} is satisfied by the cryptopuzzles in Bitcoin because (a) cryptopuzzles cannot be solved by any means other than brute force search, i.e., going through the entire space of possible solutions and checking each point individually, and (b) each point in the search space has an equal probability of being a solution. 
Properties (a) and (b) of cryptopuzzles are well-known~\cite{bitcoinbook}, and result in a linear dependency on computational power, i.e. the rate at which solution candidates are checked is linearly proportional to the computational power.

Intuitively, the need for a random search over the solution space gives less powerful miners a chance to propose a block, i.e. to ``get lucky.'' The random search may start exploring a portion of the solution space already close to the task solution, allowing even a less powerful miner to solve the BT before more powerful ones do.
If \textit{Solve} implements a completely deterministic algorithm, the most powerful miner would always solve a BT before others, thus invalidating BPP~\ref{bpp:fairness}.

It is not uncommon, however, to have partially random algorithms. This is the case when the search is guided by intelligent heuristics such as Genetic Algorithms and Simulated Annealing~\cite{genetic} where not every point in the search space has an equal probability of being a solution. In such situations, a more powerful miner can compute the heuristics and narrow the search more efficiently than a resource-poor miner.
For instance, in ML, such a powerful miner will likely be faster in computing the gradient over the training dataset. 
This may, in principle, lead to a superlinear dependency on the resource, but the exact impact depends on the specific class of BTs.

A second important requirement for proposal fairness is that experienced miners, i.e., miners that have been solving BTs for an extended period of time, do not get better chances compared to new miners.
This has been formulated in \btp{rate}{c}; however, task amortizability is detrimental to fairness as well, and thus incorporated as \btp[bold]{fairness}{b}.


\section{The Problem of Replacing Cryptopuzzles with Useful Tasks}
\label{sec:problem}

In this section, we discuss the problem of replacing cryptopuzzles with useful tasks. 
We base our discussion on the properties defined in Section~\ref{sec:properties}, and on an extended architecture for blockchain systems obtained by modifying the architecture presented in Section~\ref{sec:background}.

The remainder of this section is organized as follows. Section~\ref{subsec:usefulness} discusses different definitions of usefulness and related concepts. Next, Section~\ref{subsec:generalizing-impact} discusses the challenges of solving classes of tasks other than cryptopuzzles. Section~\ref{subsec:evolve} identifies different elements in Bitcoin-based blockchain platforms that have to be modified in order to support useful computation, while Section~\ref{subsec:usefulness-impact} discusses the impact of usefulness on BTPs.
We then describe an extended architecture based on the Bitcoin model in Section~\ref{sec:extended_architecture}.

\subsection{Usefulness Definitions}
\label{subsec:usefulness}

While Section~\ref{sec:background} presents computational tasks, there is no widely accepted definition of a \textit{useful task}. 
Most of the works surveyed in Section~\ref{sec:existing_approaches}, for example, include their own definition of usefulness.
However, while the definitions differ in formulation, they broadly agree on the principle of a task solution having value external to the blockchain.
An external party may be interested in solving one particular problem instance from a class of tasks. For example, a traveling agency may be interested in the TSP solution for a specific set of cities in which the agency operates. We call \textit{supplier} the external party that provides the task consisting of one specific problem instance to be solved. We say that this type of usefulness is \textit{strong} and is defined as follows.

\begin{definition}[Strong usefulness]
\label{def:strong_u}

There is interest external to the blockchain system in solving a specific problem instance from a class of tasks.

\end{definition}

Interest may be defined in different ways. One possibility is by financial means: if someone is willing to pay for the solution of a task, then the task is considered useful.
On the other hand, there may be interest without payments. Consider, for instance, the many BOINC-based~\cite{boinc} volunteer computing projects.
More broadly, a party may be interested in solving any problem instance within a given class of tasks.
For example, suppose a scientist is interested in having a large dataset of TSP solutions for her research. 
We say that this type of usefulness is \textit{weak} and is defined as follows.

\begin{definition}[Weak usefulness]
\label{def:weak_u}

There is interest external to the blockchain system in solving any problem instance within a class of tasks.

\end{definition}

Some proposals~\cite{primecoin, carr} argue about the utility of weak usefulness. 
It is also possible to define an intermediate degree of usefulness between weak and strong. For example, a supplier may have constraints on the instance properties, e.g. TSP solutions for any random graph of exactly 1000 vertices. 

In the case of weak usefulness, a supplier is not always necessary. A blockchain system can be designed so that problem instances of pre-defined classes of tasks can always be randomly generated and used as BT. 
We claim, however, that an external party is less likely to be interested in cryptopuzzle solutions compared to other computational tasks such as TSP, protein folding, or ML. 
For this reason, it is imperative to extend the narrow scope of blockchain platforms solving cryptopuzzles to solving other classes of tasks.

\subsection{Impact of Solving General Classes of Tasks on Task Properties}
\label{subsec:generalizing-impact}

We now discuss the challenges of satisfying BTPs for general classes of tasks. As part of the system design, there must be agreement among solvers and suppliers about the class of tasks to solve. While it may theoretically be possible to devise a system that functions without a priori agreement on the class of tasks (e.g., the task may be encoded in a language that defines the objectives in addition to the input data), it complicates the design and challenges in a way non-instrumental to the contributions of this paper. Thus, we leave this case aside. 

We observe that selecting a good class of tasks is essential because every property defined in Section~\ref{sec:properties} can be violated by a poor choice of task class. For example, finding Fibonacci numbers is amortizable, which violates \btp{rate}{c}.
Matrix multiplications require the same number of operations, violating \btp{variability}{}.
Verifying that an optimization problem solution is optimal requires solving the task again, which is problematic for \btp{verif}{b}.
A deterministic solution algorithm may disincentivize solvers from adopting incoming blocks, as explained for BPP~\ref{bpp:switch} in Section~\ref{sec:security_properties}. Furthermore, a deterministic algorithm may pose a problem for \btp{switch}{c} because a miner switching tasks ``loses'' the computation done so far. This affects block switchability, as miners close to producing a solution for the old block may prefer to do so rather than switch.

The difficulty of task classes needs to balance between \btp{rate}{a} and \btp{timeliness}{c}.
Realistically, the choice is between small-size instances of NP-hard classes and large-size instances of computational problems in P. Arguably, it is harder to adjust the difficulty of the former so that the latter may be better from the standpoint of \btp{adj}{} and \btp{adj2}{}.

Compared to cryptopuzzles, typical classes of useful tasks stem from real-world problems, and modeling such problems as computational tasks results in much higher data storage requirements for the input data and solutions.
For instance, the input to a TSP is a weighted graph, and the solution is a list of all vertices in the graph. The size of input and solution data thus grows proportionally to the size of the graph. This might be a practical issue if the tasks are stored in blocks. Since, in addition, tasks must be hard enough, the aggregated size of several TSP instances may result in a non-negligible storage space.
ML is another good example of this issue since training data may be in the order of gigabytes, if not terabytes. 
Off-chain storage solutions tackle the problem; however, in that case, the data must be fetched first. 
This impacts the required bandwidth and all of the timeliness properties, such as task generation time, solution verification time, and the actual time to solve the task. In Section~\ref{sec:classes_of_tasks}, we discuss how specific classes of tasks satisfy each BTP.

\subsection{Usefulness in Bitcoin-based Blockchain Platforms}
\label{subsec:evolve}

Adding support for usefulness to Bitcoin-based blockchain platforms relies on the design of three new elements:
(i) {\em task supply model}, i.e. how task instances are provided to the system; (ii) {\em task selection policies}, i.e. how to select the next task instances that are to be solved at any given moment; and (iii) {\em incentive model} for solving useful tasks.
This work focuses primarily on property analysis without providing a comprehensive analysis of the task supply model, selection policies, and incentive models.
Nevertheless, we discuss possible directions below. Furthermore, Sections~\ref{sec:existing_approaches} and~\ref{sec:related_work} describe different proposals found in the literature.

Regardless of how tasks are provided to the system, the system maintains a Task Supply State (TSS), which consists of a set of tasks that need to be solved or seeds used to generate the next task to solve. 
Similarly to the block context, the TSS is determined by the current state of the chain and, thus, consistent across all nodes. 
Under strong usefulness, a supplier provides the tasks to solve. In the case of weak usefulness, tasks may be generated by the system itself according to a pre-agreed algorithm, which renders a supplier unnecessary. Alternatively, suppliers may be part of the weakly useful system and may determine classes of tasks to solve. Regardless of the type of usefulness or the existence of suppliers, sources of useful tasks may be classified as internal or external to the system.

A \textit{task selection policy} determines how a node selects which task from the TSS to use as a BT for the next block. For example, a node may, in an uncoordinated fashion, freely choose any available task to solve. Another approach is to force all nodes to coordinate and attempt to solve the same task at a specific block height. 

The third additional element is the \textit{incentive model}. In a system without a supplier, it may be possible to retain the incentive model of Bitcoin based on transaction fees and a block reward. This possibility remains an open question, however, because solving useful tasks may require a different set of resources (e.g. in terms of hardware) and because a superlinear dependency on the number of node's resources (see \btp[noname]{fairness}{a}) may disincentivize less resource-rich nodes from joining the system, thus requiring additional incentives. 

Once a supplier is introduced, however, proposing a task must be subject to a fee. Such a fee not only makes sense from an economic perspective but is also required for system security and to mitigate DoS attacks. 
In fact, consider a simple incentive model consisting of modifying Bitcoin's rewards so that, all other things being equal, the coinbase is funded by the supplier's task proposal fee.
In this scenario, a malicious supplier can easily propose an already-solved task and reap the transaction fees by colluding with a miner. Clearly, this is not desired.

\subsection{Impact of Usefulness on Block Task Properties}
\label{subsec:usefulness-impact}

Replacing cryptopuzzles with useful tasks creates friction with several BTPs. In the following, we describe those issues at an intuitive level. Section~\ref{sec:classes_of_tasks} goes into the details for three selected classes of tasks.

\subsubsection{Impact on the Safety and Security Properties}
\label{sec:impact_safety}

While upholding \btp{rate}{a} primarily depends on the choice of the class of tasks, the characteristics of specific instances still play an important role. In particular, some instances may adversely affect \btp[noname]{rate}{a}. For example, a TSP where cities lay in a simple geometric form such as a line or a circle is easier to solve than the general case.
While submitted tasks can be tested against simple agreed-upon validity rules, it is infeasible to eliminate all simple instances in general.
Since untrusted suppliers may submit arbitrary task instances, this becomes a challenge for \btp[noname]{rate}{a}. 
This challenge is exacerbated by the fact that a supplier may have a monetary incentive to collude with a solver, as described earlier.
This effectively results in a {\em pre-computation attack} where the solver wins the race, gets the block reward, and splits the earnings with the supplier.

Furthermore, adjusting the difficulty requires an efficient mechanism to assess the hardness of an instance. 
Even when such a mechanism exists, adjusting the difficulty may not be feasible without violating strong usefulness. For example, adding cities to or removing cities from a TSP instance may, in principle, help adjust hardness. Yet, the resulting instance would be different.
Therefore, strong usefulness poses a challenge for \btp{adj}{} and \btp{adj2}{}.

Trivial approaches to replace cryptopuzzles with useful tasks present a clear trade-off between \btp{rate}{b} and usefulness as well. On the one hand, \btp[noname]{rate}{b} stipulates that BTs must be context-dependent.
On the other hand, usefulness implies that there is an external interest in a specific BT solution. Modifying the BT due to a context change invalidates strong usefulness in this case.
Nevertheless, non-trivial approaches may succeed in replacing cryptopuzzles with a class of useful tasks while retaining both \btp{rate}{b} and usefulness.
Section \ref{sec:proofs_of_useful_work} describes one example in detail.

\subsubsection{Impact on the Liveness Properties} 

\btp{timeliness}{b} stipulates that BTs should not be too hard to solve.
However, suppliers may submit instances that are too hard to solve, mounting a DoS attack. The same challenges described for \btp[noname]{rate}{a} apply to \btp[noname]{timeliness}{b}.

Another issue to consider in practice is that there may be no instances available in the TSS. In such case, \btp{verif}{a}[timeliness][a] does not hold true, and the system cannot progress.
In a more extreme yet plausible scenario, an empty TSS leads to an inescapable deadlock if the TSS is updated through block-included transactions. 
In Bitcoin, on the other hand, BTs are randomly generated based only on the block context. Thus, Bitcoin miners have access to an inexhaustible supply of BT. 
 
Usefulness requires ``external interest'' in the solution of a task, limiting the possible BTs -- suppliers may be interested in a subset of all possible BTs. Mechanisms for supplying and/or restricting the BTs being solved by solvers are thus necessary in order to update the TSS.

\subsubsection{Impact on the Decentralization Properties}

As described in Section~\ref{sec:decentralizations_properties}, any class of tasks whose best-known solution algorithm is deterministic is detrimental to \btp{fairness}{a}: it results in the most powerful solver being able always to solve BTs before others. Algorithms for solving useful tasks often consist of searches that get closer to the solution as they explore the solution space, giving faster solvers an advantage. In an extreme case, for instance, a few fast solvers may solve all tasks faster than all other solvers in the network, which may centralize block proposals, even though their computing power may be only slightly higher.

\begin{figure}
    \centering

    \begin{minipage}{.5\textwidth}
      \centering
      \includegraphics[width=.7\linewidth]{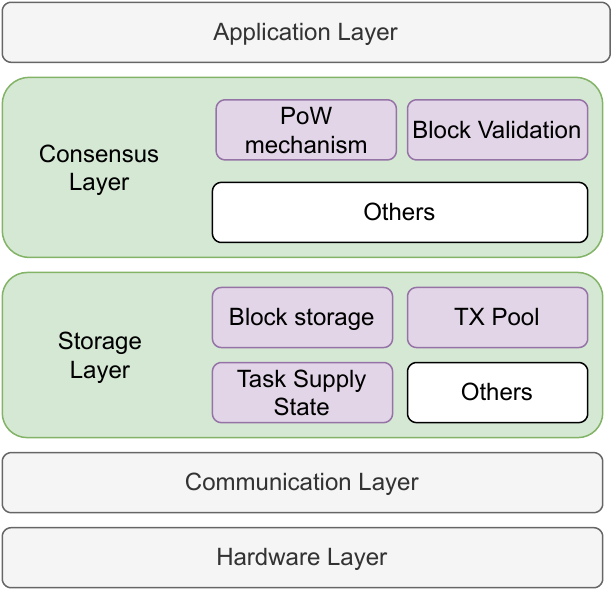}
      \captionof{figure}{Extended system architecture. Modified and newly introduced components are in a different color.}
    \label{fig:evolved_blockchain_layers}
    \end{minipage}%
    \begin{minipage}{.5\textwidth}
      \centering
      \includegraphics[width=.9\linewidth]{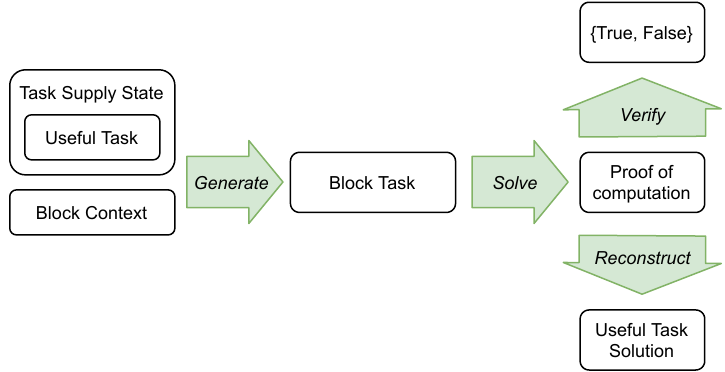}
      \captionof{figure}{Extended interface for useful tasks}
    \label{fig:evolved_interface}
    \end{minipage}

\end{figure}

\subsection{Extended Architecture and Interface to Support Useful Tasks}
\label{sec:extended_architecture}

The Bitcoin architecture introduced in Section~\ref{sec:background} does not support solving classes of tasks other than cryptopuzzles, nor does it include the design elements identified in Section~\ref{subsec:evolve}. 
Hence, we introduce an extended architecture that solves these two deficiencies. The extension is based on fundamental design considerations as well as on the study of the existing approaches described in Section~\ref{sec:related_work}.
The resulting architecture is used to discuss selected classes of useful tasks in Section~\ref{sec:classes_of_tasks} and proposed systems in Section~\ref{sec:existing_approaches}.
The roles of \textit{clients} and \textit{validators} remain the same as in Bitcoin; miners, however, are renamed to \textit{solvers} so to reflect the new responsibility: solving tasks. The role of \textit{supplier}, as already extensively described in Section~\ref{subsec:usefulness}, is introduced as well.

We start by expanding the scope of several system components, as shown in Figure~\ref{fig:evolved_blockchain_layers}. 
Transactions in the TX Pool now include all metadata necessary to support the new task supply models. Therefore the modified TX Pool supports the storage of new transaction types to, for example, advertise a task. 
The TSS introduced in Section~\ref{subsec:evolve} constitutes a separate component.
The Block storage stores all proposed blocks and the relative metadata. In practice, large tasks may lead to blocks of large sizes. In those cases, storing a portion of this data off-chain may be necessary.

The Block validation component contains the extended primitives to check the validity of an incoming block. Similarly, the PoW mechanism component implements all primitives related to computing the solution of useful tasks.
We are now ready to describe an extended interface to support the modified system architecture. 

\begin{table}[t]
\newcolumntype{x}[1]{>{\centering\arraybackslash\hspace{0pt}}m{#1}}
\scriptsize
\centering
\caption{Extended Block Task Interface}
\label{table:extended_interface}
\begin{tabular}{cx{1.5in}x{0.8in}x{1.4in}x{0.8in}}
& \textbf{Generate} & \textbf{Solve} & \textbf{Verify} & \textbf{Reconstruct} \\ 
\toprule
\textbf{Input} & Task Supply State, Task Selection Policy, Block Context & Block Task & 
Block Task, Proof of Computation & Proof of Computation  \\  
\midrule
\textbf{Output} & Block Task & Proof of Computation & Boolean & Task Solution  \\ 
\midrule
\textbf{Descr.} & 
    Generates a Block Task based on the block content, the available tasks, and the task selection policies. & 
    Solve the Block Task and outputs a Proof of Computation. & 
    Verify that the Proof of Computation is a valid solution to the Block Task. &
    Extract the BT solution from the Proof of Computation. \\
\toprule
\end{tabular}
\end{table}

Compared to the base interface, \textit{Generate} now takes as input the TSS in addition to the block context. 
Solvers generate the BT from one of the currently available instances in the TSS, chosen according to the task selection policy.
The system may enforce a specific task selection policy, e.g., the earliest unsolved instance included in the chain, or leave the selection up to the individual miner.
A simple implementation of \textit{Generate} may output a BT from the TSS without any modifications. A more elaborate construction may, e.g., transform the coordinate reference system~\cite{crs} of a TSP instance or, more generally, transform the task domain. The idea behind such a transformation is that the derived task domain may lend itself better to satisfying the BT properties. In such cases, a reverse transformation has to be applied to the solution produced by \textit{Solve} in order to reconstruct a solution to the original task, as given by the supplier.
The extended interface supports those cases.

Some of the works analyzed in Section~\ref{sec:existing_approaches} use a specific BT generation algorithm to derive the BT out of the context. Other works opt for a task prioritization algorithm and tackle \btp[noname]{rate}{b} with other means, such as shuffling of tasks (see Section~\ref{sec:proof_of_exercise}).

The semantic and input/output of the extended \textit{Solve} and \textit{Verify} remain equivalent to those in the base interface. 
We note that most of the works analyzed in Section~\ref{sec:existing_approaches} do not assume any constraint on the \textit{Solve}, which implies leaving the freedom to solvers to select any solution algorithm.


When a transformation is performed by \textit{Generate}, \textit{Reconstruct} is applied to the PoC produced by \textit{Solve} in order to obtain the solution of the original task.
Figure~\ref{fig:evolved_interface} shows the interactions between different operations of the extended interface.

\section{Selected Classes of Useful Tasks}
\label{sec:classes_of_tasks}

This section surveys some of the challenges in replacing cryptopuzzles with useful tasks. The challenge discussion is based on the property framework defined in Section~\ref{sec:properties}.
We start by identifying the general trade-offs related to optimization and decision tasks. Then, we discuss salient aspects of BTP support of the three classes of useful tasks introduced in Section~\ref{sec:background}, namely k-OV, TSP, and ML.
They represent the most common classes of tasks that have been considered in the literature, thus providing meaningful insights into the practical challenges of designing useful PoW systems. 
We have found that all these three classes satisfy \btp{verif}{a}[timeliness][a] and \btp{switch}{b}.
On the other hand, for the reasons illustrated in Section~\ref{sec:impact_safety}, \btp{rate}{b} and \btp{switch}{c} are not upheld by any of the classes.
The analysis is summarized in Table~\ref{table:table2}.

\subsection{Decision and Optimization Tasks}
\label{sec:opt_and_decision}

Solving optimization problems means finding an optimal solution. However, verifying that a solution is optimal is typically as difficult as finding an optimal solution in the first place~\cite{nphardverif}.
This applies in practice regardless of the advances in the theory of combinatorial optimization and in Verifiable Computing (VC). VC provides proof of computational integrity without any assumptions on hardware or failures~\cite{vcomp5}.\footnote{For further details on VC, we refer the interested reader to~\cite{vcomp1,vcomp2,vcomp3,vcomp4}.} A VC proof of a Concorde execution would be, in theory, sufficient to verify that the TSP solution is optimal.
However, the time overhead to create an efficiently verifiable proof for arbitrarily complex types of operations is currently prohibitive. For instance, a state-of-the-art algorithm proving a simple matrix multiplication of size 512 takes an order of minutes to execute on commodity hardware as observed in~\cite{vcomp5}. This is significantly longer than performing the multiplication itself but it can be verified faster.
In summary, as of today, achieving both \btp{timeliness}{c} and \btp{verif}{b} at the same time remains an open challenge for optimization tasks.

A possible relaxation for optimization problems is to accept suboptimal solutions that lie within a fixed margin from the optimal. 
In some cases, it is possible to efficiently estimate a bound on the optimal value of the instance. For example, the Christofides algorithm~\cite{christofides} is an approximation algorithm for TSP, which is efficient and proven to provide a solution at most 3/2 worse than the optimal solution~\cite{christofides}. 
It would be enticing to use this method to derive a validity threshold for optimization problems. 
The challenge, however, is that approximation algorithms like Christofides do not reliably produce solutions at a fixed factor i.e. they may compute a solution much better than 3/2 of the optimal. 
Consider the case in which the Christofides algorithm incidentally computed an optimal solution. Clearly, it would be impossible to produce a better solution than that, had this been the scheme to derive the margin.
In this situation, solvers cannot produce a valid solution, thus violating \btp{timeliness}{a}. 
Because of those issues, especially with regard to \btp[noname]{verif}{b}, the use of optimization problems as cryptopuzzle replacements leaves some important open problems.

On the other hand, decision problems are verifiable in polynomial time. However, the main challenge for decision problems is determining a suitable threshold: a too-pessimistic threshold makes the task easy to solve, violating \btp{rate}{a}. In contrast, a too-optimistic threshold may take a long time to produce a valid solution, violating \btp{timeliness}{c}. In the worst case, the threshold might be so high (or low) that there exists no feasible solution for the BT, which violates \btp{timeliness}{b}. 
This applies to TSP and DL tasks discussed respectively in Section~\ref{sec:tsp_tasks} and Section~\ref{sec:dl_tasks}.
Thus, the choice of the threshold and how it is calculated (or supplied) is central to the design of the solution.


\begin{table*}
\tiny 
\centering
\caption{Block Task Properties for different classes of tasks}
\label{table:table2}

\begin{tabular}{p{0.1cm}p{1.5cm}p{1cm}|p{3cm}|p{3cm}|p{3cm}} 

\toprule
     \bf {\specialcell[t]{}} & 
     \bf {\specialcell[t]{BTP}} &
     \bf {\specialcell[t]{Interface}} &
     \bf {\specialcell[t]{k-OV}} &
     \bf {\specialcell[t]{TSP}} &
     \bf {\specialcell[t]{DL}}\\
\toprule

& 
{BT hardness}  & 
{\specialcell[c]{\textit{Generate}}} &
Polynomial complexity requires infeasible large-scale input requirements &
NP-complete but vulnerable to simple special cases &
Dependent on the interplay between validity threshold, training data, and NN parameters \\ 

\cmidrule(lr){2-6}

& {{Ctx sensitivity}}  & 
{{\textit{Generate}}} &
\multicolumn{3}{c}{Trade-off with strong usefulness} 

\\ 

\cmidrule(lr){2-6}

& {{Adjustable \newline lower \newline threshold}}  & 
{{\textit{Generate}}} &
\multicolumn{2}{c|}{Tasks can be batched into harder BTs} &
Function of the validity threshold and training data. Opportunities to investigate the impact of NN parameters \\ 

\cmidrule(lr){2-6}

\multirow{6}{*}{\rotatebox{90}{Safety}}
& 
{{No reduction \newline in solvability}}  & 
{{\textit{Generate}}} &
k-OV BTs always admit a valid solution &
TSP tasks always admit a valid solution &
Dependent on the interplay between validity threshold, training data, and NN parameters. \\ 

\cmidrule(lr){2-6}

& {\specialcell[]{Negligible BT\\ generation time}}  & 
{\it \specialcell{\textit{Generate}}} &
\multicolumn{3}{c}{BTs are generated in constant time}
\\ 

\cmidrule(lr){2-6}

& {\specialcell[]{Non-amortizability}}  & 
{\specialcell{\textit{Solve}}} &
Conjectured to be difficult to amortize effectively &
Not explored in any related work & 
Invalidated by transfer learning and similar techniques \\

\cmidrule(lr){2-6}

&
{{Non-zero \newline variability \newline across BTs}}  & 
{{\textit{Solve}}} &
Commonly used solution algorithms are characterized by low variability &
Variability orders of magnitude smaller compared to cryptopuzzles. Likely trade-off with \textit{BT hardness} &
Unclear whether sufficient due to the multitude of SGD variants \\

\cmidrule(lr){2-6}

& 
{\specialcell[]{No increase in\\solution time}}  & 
{\it \specialcell{\textit{Solve}}} &
\multicolumn{3}{c}{Not satisfied} 
\\ 

\cmidrule(lr){2-6}

& {\specialcell[]{BT verification \\ soundness}}  & 
{\specialcell{\textit{Verify}}} & 
Set inclusion is verifiable in constant time &
Trade-off with \btp{verif}{b} &
Satisfied by the threshold-based approach \\

\midrule

& 
{\specialcell[]{BT generation\\efficiency}}  & 
{\it \specialcell{\textit{Generate}}} &
\multicolumn{3}{c}{BTs are generated in constant time}
\\ 

\cmidrule(lr){2-6}

& 
{\specialcell[]{BT solvability}}  & 
\textit{Generate} & 
\multicolumn{2}{c|}{BTs always admit a valid solution} &
Dependent on the interplay between validity threshold, training data and NN parameters. \\

\cmidrule(lr){2-6}

\multirow{5}{*}{\rotatebox{90}{Liveness}} & 
{{Adjustable \newline upper \newline threshold}}  & 
{{\textit{Generate}}} &
Tasks can be partitioned into easier BTs &
Trade-off with strong usefulness &
Function of the validity threshold and training data. Opportunities to investigate the impact of NN parameters \\ 

\cmidrule(lr){2-6}

 & 
{\specialcell[]{BT tractability}}  & 
{\specialcell{\textit{Solve}}} &
Achievable by partitioning BTs into easier sub-tasks &
Possibly vulnerable to hard special cases targeting specific algorithms &
Dependent on the interplay between validity threshold, training data and NN parameters \\

\cmidrule(lr){2-6}

& 
{\specialcell[]{BT verification\\efficiency}}  & 
{\specialcell{\textit{Verify}}} &
Trade-off with \btp{compl}{} &
Trade-off with \btp{sound}{} and \btp{compl}{} &
BTs are verified in polynomial time. Bandwidth may hinder efficiency \\

\cmidrule(lr){2-6}

& 
{{BT verification \newline completeness}}  & 
{{\textit{Verify}}} &
\multicolumn{2}{c|}{Trade-off with \btp{verif}{b}} &
Satisfied by the threshold-based approach \\

\midrule

\rotatebox{90}{Dctr.}  
& 
{\specialcell[]{No superlinear\\ dependency on\\ the resources}}  & 
{\specialcell{\textit{Solve}}} &
\multicolumn{3}{c}{Conjectured as not satisfied} \\

\toprule

\end{tabular}
\end{table*}

\subsection{Characterization of k-OV as a Block Task}
\label{sec:char_kov}

In the following, we characterize a blockchain based on k-OV tasks where \textit{Solve} determines which vectors are orthogonal by solving Equation~\ref{eq:ov} and the PoC contains OV sets.
While we focus on the case of $k=2$, the characterization also applies to other values of $k$.

A straightforward implementation of Equation~\ref{eq:ov} computes $n^2 d$ multiplications.
Therefore, given a computing power $c$ measured in operations per second and a time budget $t$, assuming square matrices such that $d=n$, the upper bound on the size of the matrices $n$ solved within the time budget is $(ct)^{1/3}$.
As mentioned in Section~\ref{subsec:tasks}, even the best-known algorithm iterates over all vector permutations. 
However, OV complexity is insufficient to satisfy \btp[noname]{rate}{a} due to being polynomial.
To illustrate the point, we perform a back-of-the-envelope calculation. 
The entire calculation can be found in Appendix~\ref{appendix:derivation_matrix_size} but our assessment is that a mining pool of 400 PFLOPS requires matrices of dimension $n \approx 6.2\cdot10^{6}$ 
to solve a k-OV instance in 600 seconds. 
If every vector dimension requires 1 bit of representation, each matrix of that size ($n^2$) would occupy 4,8 TB. 
Since every block must include a BT, we conclude that the space requirement of a k-OV secured blockchain is impractical.

All k-OV BTs admit a valid solution, as required by \btp{timeliness}{b} and \btp{switch}{a} because it is always possible to produce the list of OVs given enough time. A task can easily be batched with another task or decomposed into sub-problems by partitioning the input matrices, thus satisfying \btp[noname]{timeliness}{b}. Task batching facilitates \btp{adj}{}, while task partitioning helps with \btp{adj2}{} and \btp{timeliness}{c}.

In theory, k-OV is trivially amortizable if different BTs include copies of the same vector, which is detrimental for \btp{rate}{c}. In practice, the effectiveness of amortization over multiple BTs is constrained by the fact that computing multiplications and sums is a fairly cheap operation whose complexity mostly comes from the large input scale, as explained above. 

On the other hand, a limit of k-OV is that solution algorithms are deterministic and characterized by low variability. While likely non-zero, thanks to hardware heterogeneity, this may be insufficient for \btp{variability}{}. 
In addition, as anticipated in the space requirement estimates, solvers with higher FLOPS are likely to solve k-OV BT faster than other solvers due to such a deterministic solution search. This is clearly detrimental for \btp{fairness}{a}. 

\btp{sound}{} is satisfied and efficient to compute because verifying that a vector in the PoC is present in the input instance takes constant time.
\btp{compl}{}, on the other hand, presents a trade-off with \btp{verif}{b} since a PoC may not include all the existing OV. Therefore, the only way to verify the PoC is to re-execute the calculation at the expense of \btp[noname]{verif}{b}.

\subsection{Characterization of TSP as a Block Task}
\label{sec:tsp_tasks}

We now characterize a blockchain based on TSP tasks. An instance in the TSS consists of an arbitrary set of cities.
\textit{Solve} finds a valid permutation of cities which minimizes the cost as expressed by Equation~\ref{eq:tsp}. Such permutation represents the PoC, and solvers are free to choose their preferred solution algorithm.

As mentioned in Section~\ref{subsec:usefulness-impact}, simple tasks jeopardize \btp{rate}{a}. Despite TSP being NP-complete, it is extremely fast to solve small instances: it takes Concorde 10 seconds on average to find the optimal solution to TSPLIB instances with fewer than 1000 cities~\cite{hard_tsp}. Heuristic algorithms are even faster. Enforcing a minimum size on the instances, e.g. to be above 5000 cities, does not solve the issue due to the abundance of easy-to-solve special cases. 
Similarly, there exist particularly difficult instances that pose a problem a problem for \btp{timeliness}{c}. For example, the authors of~\cite{hard_tsp} present a family of instances for which Concorde takes $10^6$ longer execution times compared to instances of similar size. 
At any rate, to the best of our knowledge, hard instances have been devised for individual algorithms; hence, using alternative solution algorithms should mitigate the problem.

Unlike k-OV tasks, TSP tasks cannot be partitioned. Hence, \btp{adj2}{} poses a challenge, as previously discussed in Section~\ref{sec:impact_safety}.
\btp{adj}{}, on the other hand, may still be addressed by batching multiple BTs together.
Since a solution to the optimization task is guaranteed to exist, \btp{timeliness}{b} and \btp{switch}{a} are both satisfied.

The authors of~\cite{hard_tsp} show that multiple executions of Concorde on TSPLIB instances have different runtimes, spanning roughly one order of magnitude and centered on the average runtime. For example, an instance solved after 50 seconds on average has executions that terminate after 10 as well as 100 seconds. This result is fairly stable across the benchmarked instances of different sizes and seems promising for the support of \btp{variability}{}. However, the desired degree of variability is not precisely known: instances up to 200 cities terminate in a fraction of a second, and variability of a few seconds may be insufficient to support \btp[noname]{variability}{}. 
In Bitcoin, assuming the current level of cryptopuzzle difficulty, an ASIC computing $10^{14}$ hashes/s, such as the one described previously, has an expected time of 1157 days to propose a block and a standard deviation of $10^4$ days, based on the formula in~\cite{bitcoinvariance}. 
Therefore, to equal Bitcoin's variability, a TSP BT should terminate in about 1000 days, which is at odds with \btp[noname]{rate}{a}; faster executions may not have the required variability.

To the best of our knowledge, the question of whether and to what extent TSP tasks support \btp{rate}{c}[fairness][b] has not been explored in any related work and remains open. 

\btp{verif}{b} is a challenge for optimization tasks, as discussed in Section~\ref{sec:opt_and_decision}. Inefficient verification impacts \btp{sound}{} and \btp{compl}{} since the decision to accept or reject a solution claimed as optimal requires its computation in the first place.

We note that many state-of-the-art TSP solution algorithms follow stochastic approaches.
It is, however, unclear to which degree \btp{fairness}{a} is satisfied. The analysis is algorithm-specific, and so far, this metric has not been of interest in the analysis of algorithms, which has rather focused on other performance-related metrics. 
We conjecture that most solution algorithms for TSP do not satisfy \btp{fairness}{a}.

\subsection{Characterization of DL as a potential area for Block Tasks}
\label{sec:dl_tasks}

DL is a promising candidate as a class of tasks because it presents a high degree of customization. For instance, DNN may have varying layers, arrangement of neurons, activation functions, etc. 
Those parameters can be constrained (e.g., the number of neurons must not exceed a specific threshold) and are efficiently verifiable by inspecting the DNN. 
Nonetheless, other parameters, such as those related to training (number of training batches, type of SGD, and so on) are not efficiently verifiable by direct inspection. 
In the following, we characterize blockchain based on DL as BTs.

The TSS comprises supplier-provided datasets and corresponding error thresholds thereafter referred to as validity thresholds.
\textit{Solve} updates the set of weights according to Equation~\ref{eq:ml2}, thus minimizing the error in Equation~\ref{eq:ml}. 
The resulting DNN is used as PoC. 
Note that 
in the ML community, the error value on training data is a proxy goal for achieving good generalization performance with previously unseen data. Zero error on training data is usually a sign of overfitting, and many mechanisms, such as regularization techniques, have been developed to avoid that~\cite{regtech}. Thus, in the following, we consider a solution valid if it is above a certain arbitrary non-zero error threshold. However, the error should not be too big either. \textit{Verify} checks that the PoC satisfies such error requirements. 
Using a validity threshold makes some of the observations in Section~\ref{sec:tsp_tasks} applicable to DL BTs as well. 

Additionally, there are specific constraints due to the nature of DL tasks.
For instance, a small dataset constituted by a handful of images is quickly and efficiently learned by a DNN, thus invalidating \btp{rate}{a}. Hence, the dataset should have sufficient variety, complexity, and size.
Similarly, it has been shown that sufficiently over-specified DNNs (i.e. DNNs that are larger than needed in relation to the task at hand) are easy to train because global optima are ubiquitous and generally computationally easy to find~\cite{livni}.
Therefore, in the general case, \btp[noname]{rate}{a} depends on a non-trivial interplay between (a) validity threshold, (b) training data, and (c) DNN parameters. 
These parameters also impact \btp{switch}{a}, \btp{timeliness}{c}, and even \btp{timeliness}{b} since a poor choice may result in the absence of a solution.
We believe that calibrating these parameters effectively and on a per-instance basis remains an open challenge.

On the positive side, these parameters result in a high degree of customization inherent to DL tasks, which facilitates the implementation of \btp{adj}{} and \btp{adj2}{}.
Tinkering with the training data invalidates strong usefulness. Depending on the external interest, altering the validity threshold may also invalidate strong usefulness.
On the other hand, adapting the DNN parameters may be a viable strategy due to the ability to characterize their impact on the training time~\cite{predicting_ml}. However, additional work is needed to generalize across different DNN, data sizes and hardware configurations.
We regard this as a research gap and describe it in Section~\ref{sec:outlook}.

In ML, it is a common practice, known as transfer learning, to train a DNN on a dataset and apply the trained DNN to a different but related problem to reduce the training time~\cite{transfer_learning}. 
Transfer learning invalidates \btp{rate}{c}[fairness][b] by amortizing the training time with past computation. Therefore, a blockchain replacing cryptopuzzles with DL tasks should neutralize the training time reduction due to transfer learning, which is instead a desideratum in usual ML applications. 

\btp{variability}{} depends on the training algorithm.
SGD, along with its variants, is a popular choice for optimizing DNN. SDG is characterized by significant variability.
However, some works have criticized SGD for its slow asymptotic convergence due to its stochastic nature~\cite{sgd_variance} and proposed modifications to reduce the variance~\cite{sgd_variance, sgd_linear_convergence}.
Therefore, it remains unclear whether the cutting-edge algorithms exhibit sufficient variability.

\textit{Verify} may require the transfer and processing of large amounts of data, potentially invalidating \btp{verif}{b}. However, once the necessary data is available locally, \textit{Verify} takes polynomial time by requiring a forward pass for each data point.
In addition, the threshold-based approach guarantees \btp{sound}{} and \btp{compl}{}.

It is well-known that increasing the number of GPUs dedicated to ML training shortens the training time, in some cases up to two orders of magnitude~\cite{bert}.
In addition, the authors of~\cite{mlfairness} show that memory bandwidth is the second-order bottleneck in training time after GPU computing power. 
Therefore, miners that invest heavily in more GPUs with higher memory bandwidth are measurably faster to \textit{Solve} a BT. Following this observation, we conclude that \btp{fairness}{a} does not hold for DL and that the most powerful miners are likely to propose more than their fair share of blocks.

\section{Existing proposals for supporting Proof of Useful Work}
\label{sec:existing_approaches}

Several system designs have been proposed to solve the problem of replacing cryptopuzzles with useful tasks.
This section analyzes a representative subset of these proposals, their BTP support, as well as how each proposal positions itself wrt. usefulness, supply models, task selection policies, and incentives.
In short, none of the analyzed proposals comprehensively satisfies all the BTPs. We highlight the trade-offs of each solution and conclude that, for the time being, replacing cryptopuzzles with useful tasks remains an open problem.

In the face of the challenge, some of the proposals only introduce a partial replacement: they combine cryptopuzzles with solving useful tasks in a variety of ways. For each such ``hybrid'' system, we elaborate on the balance between useful computation and solving cryptopuzzles. 

Tables~\ref{table:works_security} and~\ref{table:works_liveness} provide an analysis overview. Each cell has been marked either as \textit{positive proof}, \textit{positive conjecture}, \textit{not conjectured}, \textit{negative conjecture}, or \textit{negative proof}.
A positive (or negative) proof, denoted as \positiveproof (or \negativeproof), means that there exists formal or empirical proof to substantiate (or refute) the validity of the property.
A positive (negative) conjecture tag, denoted as \positiveconj (\negativeconj), refers to conjectures based on intuitive, logical reasoning. Some of these conjectures are found in the literature, while others are introduced in our analysis. However, these conjectures have not been proven theoretically or corroborated experimentally. 
Not conjectured (symbol \notconj) means that there is no evidence or even conjecture one way or the other, regardless of whether the property is discussed in the proposed system. 
This happens when it is unclear how to approach reasoning in the first place, or there exist opposite arguments leading to an inconclusive outcome.

By extensive and recursive literature review, we identified 36 proposed blockchain systems based on PoW in which nodes solve useful tasks as part of the block proposal process. The count includes systems that avoid solving cryptopuzzles altogether, as well as hybrid systems where the nodes solve both cryptopuzzles and useful tasks. 

In this section, we analyze those systems that provide sufficient technical detail in their description to assess BTP support and that assume the same trust and synchronization model as Bitcoin. If either of those models significantly departs from those of Bitcoin, it affects the consideration of challenges to a great extent. 
We exclude those systems from the analysis in this section, but we mention them in Section~\ref{sec:related_work}.
Some of the works studied below exhibit significant similarities. Thus, we divide the works into six affinity groups: Cryptopuzzle equivalence, Threshold-based tasks, Complementing useful tasks with staking, Domain transformation, Shuffling of tasks, and Solution search while mining. For each group, we analyze one or two systems in detail.

We found that \btp{verif}{a}[timeliness][a] and \btp{switch}{b}  are satisfied in all the systems analyzed in the current section and, therefore, do not repeat this result for each individual system. 

Table~\ref{table:works_models} summarizes the analysis of usefulness and supply models, task selection policies, and incentives of those proposals.


\begin{table*}
\tiny
\centering
\caption{Support for security and safety Block Task Properties in surveyed systems}
\label{table:works_security}

\begin{tabular}{p{1.05cm}p{3cm}p{1.25cm}p{1.5cm}p{3.3cm}p{1.7cm}} 

\toprule  

\bf{\specialcell[t]{Paper}}&
\bf{\specialcell[t]{Limited BP rate}}& 
\bf{\specialcell[t]{Non-zero BP\\variability}}&
\bf{\specialcell[t]{Adjustable lower\\threshold}}&
\bf {\specialcell[t]{Block switchability}}&
\bf {\specialcell[t]{BP verification\\soundness}} \\

\midrule 

\specialcell[]{Primecoin~\cite{primecoin}} 
& 
{\bf (a)}~\positiveconj Hardness of searching prime chains.
\newline
{\bf (b)}~\positiveproof Context depends on the prime chain origin.
\newline
{\bf (c)}~\notconj Unclear amortizability.
& 
\notconj Unclear variability across BTs.  
& 
\positiveproof The required prime chain length is adjustable up to fractional length.
&
{\bf (a)}~\positiveproof Every task has a solution.
\newline
{\bf (b)}~\positiveproof BT generation time is negligible compared to finding a solution.
\newline
{\bf (c)}~\notconj Unclear increase in solution time.
&
\positiveconj Very high probability that invalid solutions are rejected.
\\ 

\midrule 

\specialcell[]{DLchain~\cite{dlchain}}
& 
{\bf (a)}~\negativeconj Short-term targets are unilaterally decided by the supplier. 
\newline
{\bf (b)}~\positiveproof Context depends on the training seed. 
\newline
{\bf (c)}~\notconj Unclear non-amortizability for proposed DL tasks.
& 
\positiveconj Improving the accuracy up to the target is a random trial and error process. 
& 
\notconj The effectiveness of adjustability through short-term targets remains unclear. 
& 
{\bf (a)}~\negativeconj The probability that a valid solution exists may be reduced. 
\newline
{\bf (b)}~\positiveproof BT generation time is negligible compared to finding a solution.
\newline
{\bf (c)}~\positiveconj Reaching the target is a random trial and error process.
&
\positiveproof Models having unsatisfactory accuracy are always rejected by validators. \\

\midrule 

\specialcell[]{Coin.AI~\cite{coinai}} 
& 
{\bf (a)}~\notconj Unclear hardness due to underspecified validity threshold and context-free grammar properties. 
\newline
{\bf (b)}~\positiveproof Model initialization depends on block context. 
\newline
{\bf (c)}~\notconj Unclear non-amortizability for generated NN tasks.
& 
\positiveconj Variability is likely provided by heterogeneous NN training times.
& 
\negativeconj No approach is described to increase hardness in the case of simple instances.
& 
{\bf (a)}~\positiveproof Valid solutions to tasks exist thanks to the difficulty decay. 
\newline
{\bf (b)}~\positiveproof BT generation time is negligible compared to finding a solution.
\newline
{\bf (c)}~\negativeproof Changing context increases the expected time to find a solution.
&
\negativeconj Because of the difficulty decay, unsynchronized miners may accept invalid solutions. \\

\midrule 

{Hybrid \newline Mining~\cite{hybrid}} 
& 
{\bf (a)}~\negativeproof No control over the hardness of user-supplied tasks.
\newline
{\bf (b)}~\negativeproof Useful tasks are not context-sensitive. 
\newline
{\bf (c)}~\notconj Unclear non-amortizability of NP-complete tasks.
& 
\notconj Unclear BT solution time variability for useful tasks. 
& 
\notconj Adjustability of useful tasks is not discussed and remains unclear.
& 
{\bf (a)}~\notconj The probability that a valid solution exists depends on the class of tasks.
\newline
{\bf (b)}~\positiveproof BT generation time is negligible compared to finding a solution.
\newline
{\bf (c)}~\notconj The expected solution time depends on the class of tasks.
&
\notconj Not discussed. Verification soundness depends on the type of NP-complete task. \\

\midrule 

{Proofs of \newline Useful \newline Work~\cite{ball1}} 
& 
{\bf (a)}~\positiveconj BT proved hard on average. Unclear task storage requirements.
\newline
{\bf (b)}~\positiveproof Nonce-dependent context.
\newline
{\bf (c)}~\positiveproof Polynomial evaluations are proved to be not amortizable.
& 
\negativeconj Similar number of operations for same difficulty tasks.  
& 
\negativeconj Trade-off with strong usefulness.
& 
{\bf (a)}~\positiveproof Every BT has a solution.
\newline
{\bf (b)}~\positiveproof BT generation time is negligible compared to finding a solution.
\newline
{\bf (c)}~\negativeproof The expected time to propose a new block increases. 
&
\positiveproof The property is proven by the authors. \\

\midrule 

{Proof of \newline eXercise~\cite{exercise}} 
& 
{\bf (a)}~\positiveconj High dimension input size coupled with a Proof of Hardness.
\newline
{\bf (b)}~\positiveconj Assigning miners to the task depends on the context.
\newline
{\bf (c)}~\negativeconj Matrix operations are likely amortizable.
& 
\notconj Unclear variability across BTs.
& 
\positiveconj Batching of tasks to regulate the proposal generation time. 
& 
{\bf (a)}~\notconj Depends on the class of tasks.
\newline
{\bf (b)}~\positiveproof BT generation time is negligible compared to finding a solution.
\newline
{\bf (c)}~\negativeproof The expected time to propose a new block increases. 
&
\positiveconj Invalid proposals are likely to be rejected, proportionally to the number of verifications performed.
\\

\midrule 

{Proof of \newline Search~\cite{pos}} 
& 
{\bf (a)}~\positiveproof Hardness follows from cryptopuzzles. 
\newline
{\bf (b)}~\positiveproof Context sensitivity guarantees are similar to Bitcoin.
\newline
{\bf (c)}~\positiveproof Non-amortizability follows from cryptopuzzles.
& 
\positiveproof Follows from cryptopuzzles. 
& 
\positiveproof By increasing the number of leading zeros of valid blocks. 
& 
{\bf (a)}~\negativeconj Switching to a new instance may reduce solvability.
\newline
{\bf (b)}~\positiveproof BT generation time is negligible compared to finding a solution.
\newline
{\bf (c)}~\negativeconj Switching to a new instance may increase solution time.
&
\positiveproof Invalid solutions are always rejected. \\

\midrule 

\specialcell[]{Ofelimos~\cite{ofelimos}} 
& 
{\bf (a)}~\positiveconj Hardness follows from cryptopuzzles and DPLS.
\newline
{\bf (b)}~\positiveproof Cryptopuzzles provide context dependency.
\newline
{\bf (c)}~\positiveconj Non-amortizability of BT provided by cryptopuzzles.
& 
\positiveproof Variability follows from cryptopuzzles and DPLS. 
& 
\positiveconj It may be possible to increase the difficulty by adjusting $T_2$ under certain assumptions on DPLS.
& 
\textbf{(a)}~\positiveconj Solvability of new BTs is not reduced.
\newline
\textbf{(b)}~\positiveproof BT generation time is negligible compared to finding a solution.
\newline
\textbf{(c)}~\positiveconj Solution time is likely not to increase for constant values of $T_2$.
&
\positiveconj Trivial for cryptopuzzles. Derived from SNARGs for useful tasks. 

\\
\bottomrule  

\end{tabular}
\end{table*}

\begin{table*}
\tiny 
\centering
\caption{Support for liveness and decentralization Block Task Properties in surveyed systems}
\label{table:works_liveness}

\begin{tabular}{p{1.2cm}|p{1.7cm}p{3.7cm}p{1.7cm}p{1.5cm}|p{1.8cm}} 

\toprule

\bf{\specialcell[t]{Paper}}&
\bf{\specialcell[t]{BP Verification\\efficiency}}&
\bf{\specialcell[t]{BP Timeliness}}& 
\bf{\specialcell[t]{Adjustable upper\\ threshold}}&
\bf{\specialcell[t]{BP Verification\\completeness}}&
\bf {\specialcell[t]{Proposal Fairness}} \\

\midrule

\specialcell[]{Primecoin~\cite{primecoin}} 
& 
{\bf (a)}~\positiveproof See \btp[noname]{timeliness}{a}.
\newline
{\bf (b)}~\positiveproof Fermat-based primality tests are efficient.
& 
{\bf (a)}~\positiveproof BT generation is efficient.
\newline
{\bf (b)}~\positiveproof Each BT has a solution.
\newline
{\bf (c)}~\positiveconj Length of prime chain empirically chosen to be tractable.
& 
\positiveproof The required prime chain length is adjustable up to fractional length.
& 
\positiveproof Valid solutions are always accepted.
&  
{\bf (a)}~\notconj Unclear dependency on the employed resources. 
\newline
{\bf (b)}~\notconj See \btp[noname]{rate}{c}. \\

\midrule

\specialcell[]{DLchain~\cite{dlchain}}
& 
{\bf (a)}~\positiveproof See \btp[noname]{timeliness}{a}.
\newline
{\bf (b)}~\negativeconj Efficient verification in the absence of forks. Inefficient otherwise. 
& 
{\bf (a)}~\positiveproof \textit{Generate} in constant time.
\newline
{\bf (b)}~\negativeconj Short-term targets are unilaterally decided by suppliers. It may not be solvable.
\newline
{\bf (c)}~\negativeconj Short-term targets are unilaterally decided by suppliers. It may not be tractable.
& 
\notconj The effectiveness of adjustability through short-term targets remains unclear.   
& 
\positiveproof
Validators always accept models having satisfactory accuracy.
&  
{\bf (a)}~\positiveconj Improving the accuracy up to the target is a random trial and error. 
\newline
{\bf (b)}~\notconj See \btp[noname]{rate}{c}. \\

\midrule

\specialcell[]{Coin.AI~\cite{coinai}} 
& 
{\bf (a)}~\positiveproof See \btp[noname]{timeliness}{a}.
\newline
{\bf (b)}~\positiveconj Validation requires a forward pass of the neural network and a check of the hyperparameter mapping function. 
& 
{\bf (a)}~\positiveproof \textit{Generate} takes constant time.
\newline
{\bf (b)}~\positiveproof Tasks are solvable thanks to the difficulty decay.
\newline
{\bf (c)}~\notconj Unclear tractability due to underspecified validity threshold and context-free grammar properties. 
& 
\positiveconj Threshold decay makes instances easier to solve.
& 
\negativeconj Because of the difficulty decay, unsynchronized miners may reject valid solutions. 
&  
{\bf (a)}~\negativeconj
Not discussed. The probability to solve ML instances likely grows superlinearly with the resources employed.
\newline
{\bf (b)}~\notconj See \btp[noname]{rate}{c}. \\

\midrule

{Hybrid \newline Mining~\cite{hybrid}} 
& 
{\bf (a)}~\positiveproof See \btp[noname]{timeliness}{a}.
\newline
{\bf (b)}~\notconj Not discussed. Useful tasks may be inefficient to verify.
& 
{\bf (a)}~\positiveproof \textit{Generate} runs in constant time for both cryptopuzzles and useful tasks.
\newline
{\bf (b)}~\negativeconj Useful tasks may not have valid solutions.
\newline
{\bf (c)}~\negativeconj Tasks without valid solutions cannot be solved fast enough.
& 
\notconj Adjustability of useful tasks is not discussed and remains unclear.
& 
\notconj Verification completeness of tasks is not discussed.
&  
{\bf (a)}~\negativeconj Useful task solution probability is likely to grow superlinearly with the resources employed.
\newline
{\bf (b)}~\notconj See \btp[noname]{rate}{c}. \\

\midrule

{Proofs of \newline Useful \newline Work~\cite{ball1}} 
& 
{\bf (a)}~\positiveproof See \btp[noname]{timeliness}{a}.
\newline
{\bf (b)}~\positiveproof The verification efficiency is proved to be linear.  
& 
{\bf (a)}~\positiveproof \textit{Generate} runs in linear time.
\newline
{\bf (b)}~\positiveproof The polynomial has a valid solution.
\newline
{\bf (c)}~\positiveconj Tractability can be controlled via the number of variables in the polynomial.
& 
\negativeconj Trade-off with strong usefulness.
& 
\positiveproof The property is proven by the authors. 
&  
{\bf (a)}~\negativeproof The solution algorithm is deterministic, which favors centralization.
\newline
{\bf (b)}~\positiveproof See \btp[noname]{rate}{c}. \\

\midrule

{Proof of \newline eXercise~\cite{exercise}} 
& 
{\bf (a)}~\positiveproof See \btp[noname]{timeliness}{a}.
\newline
{\bf (b)}~\notconj Unclear efficiency in practice.
& 
{\bf (a)}~\positiveproof \textit{Generate} runs in constant time.
\newline
{\bf (b)}~\notconj Depends on the class of tasks.
\newline
{\bf (c)}~\positiveconj Miners are assigned sub-portions of the input matrix as tasks.
& 
\positiveconj Task hardness is reduced by dividing tasks into easier sub-problems.
& 
\positiveproof Valid solutions are always accepted.  
&  
{\bf (a)}~\negativeconj Common solution algorithms are deterministic.
\newline
{\bf (b)}~\negativeconj See \btp[noname]{rate}{c}. \\

\midrule

{Proof of \newline Search~\cite{pos}} 
& 
{\bf (a)}~\positiveproof See \btp[noname]{timeliness}{a}.
\newline
{\bf (b)}~\positiveproof Solutions are verified in constant time. 
& 
{\bf (a)}~\positiveproof \textit{Generate} runs in constant time.
\newline
{\bf (b)}~\negativeconj A small task solution space may lead to unsolvable BTs. 
\newline
{\bf (c)}~\negativeconj Tractability is guaranteed only when assuming a sufficient task solution space. 
& 
\negativeconj 
It is not possible to reduce the difficulty of unsolvable instances.
& 
\positiveproof Valid solutions are always accepted. 
&  
{\bf (a)}~\notconj Depends on the class of tasks and the solution search cost compared to cryptopuzzles.  
\newline
{\bf (b)}~\positiveproof See \btp[noname]{rate}{c}. \\

\midrule

\specialcell[]{Ofelimos~\cite{ofelimos}}
& 
{\bf (a)}~\positiveproof See \btp[noname]{timeliness}{a}.
\newline
{\bf (b)}~\notconj Assumptions on SNARGs efficiency may not be satisfied in practice.
& 
{\bf (a)}~\positiveproof \textit{Generate} in constant time.
\newline
{\bf (b)}~\positiveproof All tasks have a valid solution.
\newline
{\bf (c)}~\positiveconj Tractability follows from cryptopuzzles and DPLS.
& 
\positiveconj It may be possible to reduce the difficulty by adjusting $T_2$ under certain assumptions on DPLS.
& 
\positiveconj Trivial for cryptopuzzles. Derived from SNARGs for useful tasks.
&  
{\bf (a)}~\notconj Fairness claimed, but the proof is omitted.
\newline
{\bf (b)}~\notconj See \btp[noname]{rate}{c}. 

\\ 
\bottomrule

\end{tabular}
\end{table*}

\subsection{Attempts at cryptopuzzle equivalence}
\label{sec:primecoin}

As the high hurdle of the challenge of replacing cryptopuzzles was unfolding over the years, later proposals came up with sophisticated techniques that were trying to compensate for the property gap between cryptopuzzles and other classes of computational tasks. On the other hand, earlier proposals attempted to carefully craft a dedicated class of tasks that would be useful, and yet would possess properties similar to those of cryptopuzzles. Among those works are Primecoin~\cite{primecoin} and Gapcoin~\cite{gapcoin}. While motivating and inspiring later ideas, the crafted classes were criticized for the lack of practical applications, which limited their usefulness. Besides, not all requirements were analyzed or even known. Hence, despite being carefully crafted, the tasks do not satisfy all of the BTPs we list in Section~\ref{sec:properties}.

\textbf{Primecoin} was introduced in 2013 and is considered the first practical attempt to replace cryptopuzzles with useful computation. 
In Primecoin, solvers find long sequences of prime numbers, also known as chains. Primecoin accepts three types of chains. In the first type, called Cunningham chain~\cite{guy} of the first kind, each prime number (except the first) must be equal to twice the previous prime number minus one, for example, 6121, 12241, and 24481. In the second type, called Cunningham chain of the second kind, each prime must equal twice the previous prime plus one. The last type combines the previous two and is called bi-twin; a bi-twin chain is a chain where primes form pairs that are two units apart from each other and where the average of each pair is double the average of the previous pair. An example of bi-twin chain is 211049, 211051, 422099, and 422101.
For all three chain types, it is possible to define an \textit{origin} i.e. a number close to the first element of the prime chain. In the numerical examples provided earlier, the two origins are respectively 6120 and 211050.
Primecoin makes the work context-dependent by requiring the origin of the prime chain to be divisible by the block header hash. Therefore, miners first assemble a block proposal and then start searching for valid prime chains.
 
Primecoin uses probabilistic primality tests, namely the Fermat test~\cite{fermat} and the Euler-Lagrange-Lifchitz test~\cite{lif}, to verify the validity of a block. Those tests check whether the equality in Fermat's little theorem~\cite{hardy} holds for a set of randomly chosen values. The confidence that a number is prime increases with the number of tested values that satisfy the equality.

Generally, the longer the prime chain, the harder it is to find one.
However, since a chain longer by just one prime number may be considerably harder to find, Primecoin accepts fractional chain lengths e.g. 11.6, where the decimal part is the Fermat test remainder. Primecoin uses such a remainder to approximate a linear difficulty function.

Since the tasks are not user-supplied and the system accepts any prime chain of sufficient length, we classify such a class of tasks as weakly useful. For the same reason, there is no task selection policy. Miners allocate all their computing steps in finding a Cunningham chain of sufficient length.
Like in Bitcoin, Primecoin miners receive block rewards and transaction inclusion fees.

\btp{rate}{a} and \btp{timeliness}{c} stem from an appropriately chosen length of the prime chain, since a longer chain is harder to find. Such length is adjustable up to fractional values, supporting both \btp{adj}{} and \btp{adj2}{}.
\btp{rate}{b} is guaranteed by tying the block to the origin of a prime chain.
\btp{rate}{c}[fairness][b] is not discussed in the paper and remains unclear whether solutions are amortizable, for example, by selecting a block for which the associated origin has an already known chain of primes.
Similarly, \btp{variability}{} is not discussed in the paper, and due to the lack of analysis, it remains unclear to what extent the property is satisfied.

Probabilistic primality tests like Fermat's are efficient, \btp{verif}{b}, insure \btp{compl}{}  and provide \btp{sound}{} with very high probability, especially when repeated multiple times.

\btp{fairness}{a} is not discussed, and the extent of fulfilling it remains unclear.

From Dickson's conjecture~\cite{dickson} and Schinzel's hypothesis H~\cite{schinzel}, both widely believed to be true, it follows that for any positive number $k$ there are infinitely many Cunningham chains of length $k$. Those conjectures guarantee the existence of a valid solution; thus, Primecoin satisfies both \btp{timeliness}{b} and \btp{switch}{a}.
It remains unclear, however, whether the expected time to find a valid solution to a new BT, \btp{switch}{c}, does not increase.

\newcommand{\DLth}[0]{$\delta$\xspace}

\subsection{Attempts at replacing cryptopuzzles with threshold-based tasks}

As mentioned above, earlier attempts at cryptopuzzle equivalence drew two lines of criticism on the proposed classes of tasks: their practical utility appeared insufficient and the stated goal of equivalence in terms of satisfying all cryptopuzzle properties was not achieved. A number of later attempts focused on utilizing classes of high practical utility (such as ML); they were mindful about the properties without expecting to reach the tantalizing ideal of full equivalence. Specifically, works in these category considered threshold-based decision tasks derived from useful optimization tasks. Valid solutions were those scoring above (or below) a predefined threshold according to some optimization metric; thus thresholds were used to emulate the difficulty of cryptopuzzles. The metric is typically related to usefulness, so suppliers are directly interested in solutions scoring better than the threshold. 
As extensively discussed in Section~\ref{sec:opt_and_decision}, the choice of threshold impacts BTPs at different levels and it is challenging to make a good decision regarding the threshold value.

The works in this group are based on ML~\cite{dlchain, coinai, pole} and other optimization tasks~\cite{transportation}. In the following, we discuss two works in ML.

\textbf{DLchain}~\cite{dlchain} is a system that uses Deep Learning (DL) training as the class of tasks for a permissionless ledger. DL tasks are submitted by trusted suppliers that maintain a permissioned blockchain among themselves. The goal of this second ledger is to establish the order in which tasks need to be solved; however, DLchain does not elaborate further on the permissioned part. For example, it remains unclear how tasks and their relative order are disseminated to the solvers.
Suppliers specify a \textit{desired accuracy threshold} \DLth for every task, and DL models need to satisfy \DLth to be considered valid. 
If a task does not achieve \DLth within a system-wide deadline, the task is terminated and skipped.
The protocol introduces intermediate milestones towards \DLth, called \textit{short-term targets}. The idea is that the \DLth may take too long to achieve, so a block can be proposed as soon as the trained model satisfies an (easier) short-term target.
At any rate, the mechanism's description is qualitative, and it remains unclear how reliable the short-term target mechanism is in providing timeliness guarantees. 
Unlike suppliers, solvers form a permissionless network and are free to join and leave the system.
The solvers train the model on the data using a training seed $S$ that depends on the block header.
During training, solvers save training metadata into a list of off-chain checkpoints constituting the so-called \textit{epoch chart} and include the epoch chart hash in the block header. As we explain below, the \textit{epoch chart} is used as a reference point in the case of forks.
Upon successful block inclusion, solvers are awarded the block reward and the transaction fees.

The proposal aims to achieve strong usefulness, and we note that all computing steps are
spent on useful tasks.
However, the system critically relies on two strong assumptions. The first assumption is that task suppliers pre-train a DL model to correctly estimate and assign the value of \DLth so that improving accuracy becomes challenging and stochastic. This creates a clear computational burden on the supplier side that may deter use. The second assumption is that improving the model performance up to \DLth is useful for the supplier. We argue that this may not necessarily be the case.
Indeed, as already discussed in Section~\ref{sec:dl_tasks}, several studies show that over-optimizing the DL model to the training data, as proposed in the solution, leads to loss of generality and may underperform with new data samples, undermining the motivation about usefulness.

In the proposed scheme, \btp{rate}{a}, \btp{timeliness}{c} and \btp{timeliness}{b} depend on the task and the short-term target values, both of which are unilaterally decided by the task supplier. Suppliers are assumed to be trusted; however, they may be genuinely too optimistic or pessimistic in setting \DLth and short-term targets. Those may be outright infeasible or intractable, impacting \btp[noname]{timeliness}{b} and \btp[noname]{timeliness}{c} respectively. 
For the same reason, \btp{switch}{a} may not hold. However, DLchain counters that risk at the system level with the introduction of the system-wide deadline.

\btp{rate}{b}, on the other hand, is guaranteed due to using the training seed $S$.
A model is considered valid only when it satisfies the short-term target.
Verification is efficient in the lack of conflicts, which satisfies \btp{verif}{b}. Upon conflict, however, nodes must use the \textit{epoch chart} to verify step by step that the solution matches the chart, an extremely inefficient process. In such a situation, the winning branch is the one whose {\it epoch chart} has logged the biggest number of training iterations (epochs). 
Because of this overhead, \btp[noname]{verif}{b} is, in the worst case, as expensive as solving the task in the first place.

According to the authors, improving accuracy up to at least the value of a short-term target is a random trial and error process similar to the cryptopuzzle brute-force search. Therefore, the solution search is likely to guarantee a sufficient degree of randomness for \btp{variability}{}, \btp{fairness}{a}, and \btp{switch}{c}. 
On the other hand, fairness is not explicitly considered by the authors as a property and deserves further validation. 

Besides, the authors do not explicitly consider \btp{adj}{} and \btp{adj2}{}.
We argue that while tinkering with the short-term targets has an effect on the task difficulty, this approach is not as well-studied as the difficulty mechanism in PoW. 
Similarly, \btp{rate}{c}[fairness][b] is not discussed, and the degree of non-amortizability remains unclear for DL tasks.

Both \btp{sound}{} and \btp{compl}{} hold because models with sufficient accuracy are accepted, and the network rejects models with insufficient accuracy.
As a final remark, proposed blocks must include the model. However, the model size is non-negligible, and the space requirement will likely scale to several GB within the span of a fairly small number of appended blocks.

\textbf{Coin.AI}~\cite{coinai} replaces cryptopuzzles with the training of DNNs. Solvers can choose the training algorithm they prefer to implement \textit{Solve}, and ML models are considered valid only if they perform above a certain quality threshold. 
The threshold decays over real-time, which ensures that a block is eventually proposed.
While central to the design, Coin.AI does not explain how the quality threshold is initially set and how fast it decays, nor does it discuss the related synchronization challenges for geographically distributed nodes.
\textit{Generate} uses a surjective function to map the block hash value into an NN architecture and initial hyperparameters. This mapping function is implemented through formal context-free grammar. A modification to any block element (e.g. the block proposer) changes the input to the mapping and, consequently, the resulting BT.

Any node with sufficient funds to pay a task proposal fee can propose an instance to be solved. The instance is unaltered, so we classify the tasks under strong usefulness and note that all computing steps are spent on useful tasks.
Conversely, solvers receive rewards for proposing blocks and including transactions in a way similar to Bitcoin.
The authors discuss a few alternative options for the task selection policy, but details are not reported, and the description remains at a high level. 

Both \btp{rate}{a} and \btp{timeliness}{c} rely on the hardness assumption of the training process. While it is commonly assumed that NNs are computationally hard to train, the time to produce a solution in practice depends on the interplay between the validity threshold, the NN, and the training data as discussed in Section~\ref{sec:dl_tasks}. Coin.AI does not go into the details and properties of the context-free grammar and initial threshold selection, hence it is unclear whether \btp[noname]{rate}{a} and \btp[noname]{timeliness}{c} are satisfied.
The context-free grammar approach used to generate BTs guarantees \btp{rate}{b}.
On the other hand, it is unclear to what extent this approach prevents transfer learning, which is undesirable for \btp{rate}{c}[fairness][b], as explained in~\ref{sec:dl_tasks}. Indeed, classic transfer learning requires the layers of the pre-trained and the receiving neural network to be compatible, an occurrence that is made unlikely by context-free grammar.
On the other hand, we observe that solvers may rearrange block transactions to search for specific NN configurations.
It, therefore, remains unclear whether \btp[noname]{rate}{c}[fairness][b] is satisfied.

Coin.AI implicitly assumes that all instances are ``hard enough'' and discusses the threshold decay mechanism, without analyzing the case in which the BT may be extremely easy and quickly solved by most NN. Thus, while \btp{adj2}{} is likely satisfied through the decay mechanism, we argue that \btp{adj}{} is not.

Solvers must agree on the current time in order to calculate the current quality threshold correctly and guarantee both \btp{compl}{} and \btp{sound}{}. However, it is not uncommon in blockchain systems to experience substantial time differences. 
The ramifications of loose synchronization and related incentives, while detrimental to the blockchain properties, are not discussed in Coin.AI.
Verification is efficient, satisfying \btp[noname]{verif}{b}, since re-computing the mapping function and performing a single propagation of the dataset through the solution model are both fast operations.

\btp{variability}{} is not considered either. However, it is likely satisfied due to the heterogeneity of NN returned by \textit{Generate}.

Difficulty decay ensures that even hard instances are solvable over time, thereby guaranteeing both \btp{timeliness}{b} and \btp{switch}{a}. 
On the other hand, \btp{switch}{c} is not satisfied: upon change of context, the expected time to solve a new BT increases since miners have to abandon the currently trained NN and start anew.
The chain selection rule is not explicitly described. The description hints that better-performing models weigh more to establish the blocks belonging to the main chain. In such a case, we observe that miners may prefer to keep solving the current BT rather than to switch to a new task, and therefore, it is unclear whether miners are incentivized to switch or rather to keep improving their model.
\btp{fairness}{a} is not discussed, however, as argued in Section~\ref{sec:dl_tasks}, we believe that DL BTs do not satisfy \btp[noname]{fairness}{a}. It remains to be seen in practice to what extent small solvers are able to propose blocks in Coin.AI.

\subsection{Complementing useful tasks with a staking mechanism}

One of the challenges in replacing cryptopuzzles with useful work is estimating and controlling the difficulty of useful tasks. The systems in this group~\cite{hybrid, amar, crowdsourcing, coop-hybrid} complement solving useful tasks with an additional mechanism whose role is to contribute to controlling the overall difficulty. The exact mechanism differs across these systems but it invariably consists of PoS and staking elements, with an occasional addition of cryptopuzzles.
The staking mechanisms indeed contribute to controlling the difficulty: supplier-side deposits may help with \btp[noname]{timeliness}{b} and \btp[noname]{timeliness}{c} since it becomes the supplier's interest that the task is solvable and tractable.
Similarly, solver-side deposits may be used when committing to a solution, hence deterring task reuse and helping with \btp[noname]{rate}{b}.
It should be noted, however, that the security of collateral payment systems is, unlike Bitcoin, based on the assumption of rational attackers.

\textbf{Hybrid Mining} (HM)~\cite{hybrid} proposes a hybrid PoW system in which solvers choose whether to solve NP-complete tasks or cryptopuzzles.
Block proposers are awarded the transaction fees and either a task solution reward or the block reward, depending on the task type included in the block. The task solution reward is paid by the supplier instead of being newly minted, as happens with the block reward. Thus, the economics of HM dictate that at least half of the blocks are secured by cryptopuzzles. Hence, the fraction of useful work performed by the system is measured in blocks rather than computing steps.
HM does not explicitly specify whether the useful tasks must be decision or optimization problems. Boolean satisfiability problems (SAT) are used throughout the paper as the working example.
SAT is a decision problem, and therefore, in the following analysis, we assume this flavor of tasks.
We note that if no variable assignment exists such that the Boolean formula evaluates to true, the SAT instance is unsatisfiable, incurring all the challenges related to verifying negative claims, as discussed in Section~\ref{sec:opt_and_decision}. 

Tasks are advertised as transactions over the network.
Suppliers lock several funds in the advertisement, including the task solution reward and an advertisement fee equivalent to the block reward, to deter collusion between solvers and suppliers.
Any node can submit a task regardless of complexity as long as all fees and deposits are paid.
Usefulness is not clearly defined, but instead informally mentioned as \textit{problems of immediate practical value to their owners}.
The proposed system does not modify the useful task based on the context so that it satisfies strong usefulness based on Definition~\ref{def:strong_u}. 
To prevent stealing task solutions, HM uses a two-phase mechanism to safely disclose solutions. In the first phase, miners propose a block containing a claim that a specific task was solved. Such a block does not contain any PoW but rather a security deposit. The task solution is thereafter revealed as a blockchain transaction.
If a solver fails to disclose the useful task solution within a predefined number of blocks, the solver loses the associated security deposit. At any rate, a useful block that has been included in the chain remains part of the chain regardless of correct solution disclosure.
HM retains Bitcoin's longest chain selection rule. However, the authors admit that the difficulty of problem instances is not well-defined and is hard to compare across tasks. Therefore, the main chain in HM is decided based only on which of the competing branches has the most cryptopuzzle-secured blocks.
Since solvers are free to choose whether to solve cryptopuzzles or useful tasks, we argue that they are incentivized to solve cryptopuzzles to increase their chances of block inclusion.
In the following, 
the analysis refers mainly to the useful task construction since cryptopuzzle properties have been extensively described in Section~\ref{sec:properties} and can be easily inferred. Unless otherwise noted, a property is only satisfied at the system level if both useful tasks and cryptopuzzles satisfy it.

In the proposed construction, 
there is no guarantee on the \btp{rate}{a} of a task: suppliers may provide a series of very simple instances, or malicious solvers may have enough funds to disseminate non-solved useful blocks, causing an undue increase of the BP rate. The latter case, however, would deplete the funds of the malicious miners.
On the other hand, submitted tasks may not have a correct solution (a valid Boolean assignment to the SAT clauses that evaluates to true), jeopardizing \btp{timeliness}{b} and indirectly \btp{timeliness}{c}.
\btp{variability}{} as well as \btp{adj}{} and \btp{adj2}{} are not discussed, and while they hold for cryptopuzzles, it is unclear how arbitrary computational tasks may fare with respect to those properties. 
\btp{rate}{b} is not satisfied by useful tasks since the solution to the useful task does not depend on the context and can be solved irrespective of the state of the chain.

\btp{switch}{a} is not considered in the work and depends on the class of tasks. For example, the probability of existence of a valid solution for the new BT is reduced if the SAT instance is unsatisfiable. On the other hand, miners may opt to solve cryptopuzzles which do not reduce such probability, however, at the expense of usefulness. 
An interesting observation is that, since useful blocks do not count for the longest chain rule, the incentive for miners to switch to a newly received useful block is unclear since ignoring it yields no penalty, and extending it yields no benefit.

\btp{switch}{c} must be analyzed separately for cryptopuzzles and SAT instances.
The expected time to find a valid solution for a cryptopuzzle does not increase; on the other hand, even assuming the existence of a valid solution for a given SAT instance, the time to find a solution may be longer. Thus, \btp[noname]{switch}{c} depends on the class of tasks, and it is not satisfied in the case of SAT.

Some properties are not considered in HM. Specifically \btp{rate}{c}[fairness][b], \btp{verif}{b}, \btp{sound}{}, \btp{compl}{}. \btp[noname]{verif}{b}, \btp[noname]{sound}{}, \btp[noname]{compl}{} hold for SAT problems; however, those properties may not be guaranteed for other classes of tasks.
Regarding \btp{fairness}{a}, useful tasks such as SAT instances may lead to a superlinear probability increase. We conjecture the lack of fairness as a likely problem and raise this point as an open question requiring further analysis.

\subsection{Domain transformation}
\label{sec:proofs_of_useful_work}

The group of Domain transformation uses techniques that transform the computation of the useful task from one domain to another to improve the BTP coverage. The solution can be quickly and verifiably reconstructed (see Section~\ref{sec:extended_architecture}) from the miners' response.
This approach requires strong theoretical foundations and applies to very specific classes of tasks. Proofs of Useful Work~\cite{ball1} belongs to this group.

\textbf{Proofs of Useful Work} (PoUW) proposes a system that replaces cryptopuzzles with the evaluation of low-degree polynomials derived from instances of k-OV.
Bigger k-OV instances generate polynomials with more terms, which are harder to solve.
An updated version~\cite{ball2} of PoUW retracts the definition of usefulness and omits the design of the blockchain system; in the following, we analyze the original PoUW paper, including the system design, because we believe it is an important example worth deeper analysis.
The original usefulness definition is \say{Computational tasks can be delegated as challenges to the workers such that the solution to the delegated task can be quickly and verifiably reconstructed from the workers’ response}~\cite{ball1}. No external interest is mentioned in this definition; however, we classify PoUW as satisfying strong usefulness since a \textit{Solve} output contains the result of a specific user-supplied instance.
All computing steps are spent on useful tasks.
PoUW does not define a supply model and assumes unlimited tasks collected in a publicly accessible board. We believe this to be a strong assumption since the progress of the chain relies on the availability of those tasks. At any rate, in the following analysis, we assume that tasks are always available.
Miners select arbitrarily which task to solve among the ones on the board. 

It remains unclear what incentivizes solvers to select a bigger OV instance, thus requiring more time to solve the corresponding polynomial. In general, the incentives of the construction are not described. The k-OV instance, along with a block-dependent nonce, is used to \textit{Generate} a polynomial 
whose evaluation is proven to be hard in the average case. For a tutorial on average-case fine-grained complexity theory, refer to~\cite{average_complexity}. 

Unlike the simple design in Section~\ref{sec:char_kov}, PoUW tasks are not restrained to $k=2$. Tasks may have higher polynomial complexity, and the authors go to lengths to prove that the low-degree polynomials are hard in the average case. However, we deem \btp{rate}{a} as likely satisfied (as opposed to fully satisfied) because no observation is made about the storage requirements of the matrices which, as discussed in Section~\ref{sec:char_kov}, can be substantial.
In addition, the authors prove \btp{rate}{c}[fairness][b] for polynomials, and the block-dependent nonce in the construction supports \btp{rate}{b}. 
On the other hand, \btp{variability}{} is unlikely to be satisfied: \textit{Solve} is deterministic, and miners with similar computing power are likely to propose a block at roughly the same time.
\btp{timeliness}{b} and \btp{switch}{a} are satisfied since all k-OV tasks have a solution. \btp{timeliness}{c} is not discussed, but we surmise that \btp[noname]{timeliness}{c} can be satisfied by controlling the number of variables in the polynomial. 

Similarly, \btp{adj}{} and \btp{adj2}{} are not discussed either. Since the size and degree of the polynomial are derived from the k-OV instance, a given instance determines the hardness of the task. Based on our understanding, the derivation mechanism does not naturally accommodate hardness adjustment mechanisms, and hence, there is a trade-off with strong usefulness.

\btp{switch}{c} is not satisfied because a miner switching tasks ``loses'' the computation done so far, with the impact described in Section~\ref{sec:impact_safety}.

The authors prove \btp{sound}{} and \btp{compl}{} of their scheme.
\btp{verif}{b} is shown to have linear complexity $\tilde{O}(n)$, as a function of the number of vectors in the k-OV instance. On the other hand, in the general case, the low-degree polynomial may include many terms whose number is proportional to the k-OV instance size, leading to proofs of non-negligible size. Therefore, storage requirements may be an important practical constraint in a real deployment.
\btp{fairness}{a} is not explicitly considered in PoUW, and the solution search algorithm is deterministic: the probability of proposing a block increases superlinearly with the relative amount of resources spent. This raises concerns about the decentralization of the solution. 

\begin{table*}
\tiny 
\centering
\caption{Usefulness and supply models, task selection policies and incentives of the selected proposals.}
\label{table:works_models}

\begin{tabular}{p{1.5cm}p{2cm}p{2.8cm}p{2.8cm}p{2.8cm}} 

\toprule

\bf{\specialcell[t]{Paper}}&
\bf{\specialcell[t]{Usefulness Model}}& 
\bf{\specialcell[t]{Supply Model}}&
\bf{\specialcell[t]{Task Selection Policy}}&
\bf{\specialcell[t]{Incentives}} \\

\toprule

{Primecoin~\cite{primecoin}} 
& 
{Weak usefulness \newline All computing steps are spent on useful tasks.}
& 
Miners generate BT independently like in Bitcoin.
&
Not applicable
&
Not discussed  \\ 

\midrule

{DLchain~\cite{dlchain}} 
&
{Strong usefulness \newline All computing steps are spent on useful tasks.}
& 
Restricted to suppliers within a permissionless blockchain
&
Restricted to a single task at a time decided by the consortium of suppliers
&
Block reward and transaction fees \\ 

\midrule

{Coin.AI~\cite{coinai}} 
&
{Strong usefulness \newline All computing steps are spent on useful tasks.}
&
Any currency holder in the network can submit a task.
&
Possible Task Selection Policies are discussed at a high level.
&
Block reward and transaction fees
\\ 

\midrule

{Hybrid \newline Mining~\cite{hybrid}} 
&
{Strong usefulness \newline At most half of the appended blocks contain useful tasks.}
& 
Any node in the network can submit a task.
&
Solvers decide which tasks to solve, either a cryptopuzzle or a useful task.
&
Proposer receive transaction fees and either solution reward or coinbase reward.   
\\ 

\midrule

{Proofs of \newline Useful Work~\cite{ball1}} 
& 
{Strong usefulness \newline All computing steps are spent on useful tasks.}
& 
Not discussed
&
Arbitrary task selection from a public board
&
Not discussed  \\ 

\midrule

{Proof of \newline eXercise~\cite{exercise}} & 
{Strong usefulness \newline All computing steps are spent on useful tasks.}
& 
Any node in the network can submit a task.  
&
Solvers are assigned randomly the task to solve.
&
Deposits to disincentivize misbehaving and block reward to incentivize task solutions.  \\ 

\midrule

{{Proof of \newline Search~\cite{pos}}} 
&
{Strong usefulness \newline Unknown, highly dependent on the solution search algorithm}
& 
Any node in the network can submit a task.
&
Restricted to a list of tasks embedded in the last valid block.
&
Separate rewards for proposing a block and for the best solution to the useful task. \\ 

\midrule

{Ofelimos~\cite{ofelimos}} 
&
{Strong usefulness \newline At most half of the computing steps are spent on useful tasks.}
&
Not discussed
&
Tasks are assigned to a number of consecutive blocks. Task prioritization is based on an unspecified scheduling algorithm.
&
Rewards are given to ranking blocks and input-blocks proposers. Transaction fee assignment is similar to Bitcoin.\\

\bottomrule

\end{tabular}
\end{table*}

\subsection{Shuffling of tasks}
\label{sec:proof_of_exercise}

Altering a supplied task based on the context raises the issue of strong usefulness described in Section~\ref{subsec:usefulness-impact}. Therefore, a group of works enforces context sensitivity through the process that assigns the task to a miner rather than by modifying the task itself. This process is random but verifiable and uses an assignment service that has been referred to as shuffling.
Proof of eXercise~\cite{exercise} is the first to propose that approach, followed by~\cite{todorovic}. We analyze the former in depth.

\textbf{Proof of eXercise} (PoX) is a system in which cryptopuzzles are replaced by matrix-based computation problems, motivated by the claim that such problems span a wide range of useful real-world use cases and that they are an important abstraction for many scientific computation tasks. PoX claims to support a broad set of matrix operators, including sum, product, Schur product, etc. In the proposed system, two shuffling services match tasks and nodes in a uniformly random fashion.  
The first shuffling service assigns miners to the task to be solved, while the second service assigns the output of the \textit{Solve} function to a subset of validators. The former service is used to limit the effectiveness of collusion attacks, while the latter is to limit the
bias in verification and to improve scalability. 
We note, however, that while being central to the PoX design, shuffling services are underspecified. PoX suggests implementing such services on TOR and onion routing but does not include the discussion of challenges, practicality, and incentives. 

The paper does not provide a clear definition of usefulness. However, the proposed solution seems to target strong usefulness as in Definition~\ref{def:strong_u}. 
All computing steps are spent on matrix operations.
Any node can supply a task by storing the task instance in a highly accessible database and thereafter advertising the task in a blockchain transaction.
While task suppliers and solvers pay a deposit that is returned if they do not misbehave, block proposers are incentivized to solve the task to collect the coinbase reward. 

The proposed design assumes the existence of a routine to efficiently assess the task runtime based on the matrix sparseness and algebraic structure; such estimate is attached to the task as a \textit{Proof of Hardness} (PoH). To ensure \btp{rate}{a}, accepted tasks are those whose PoH exceeds a pre-defined threshold.  
The mechanism to satisfy \btp{rate}{b} is based on the pseudo-random task selection: if the block content is modified, the pseudo-random matching will very likely be invalidated. 
Therefore, the security of such an approach crucially relies on (1) the implementation of the shuffling service and (2) the sets of supplied tasks and solvers being large enough. The author suggests matching miners and tasks through the hash of the block header. We observe that such matching requires active miners to be registered on the blockchain. Besides, such membership must be maintained amidst node churn.
\btp{rate}{c}[fairness][b] is not discussed in the work; we argue that it is possible to amortize the solution search across instances if there is substantial overlap among them. However, the practicality of such an attempt depends on the specific class of tasks. 

Similarly, \btp{timeliness}{b} and \btp{switch}{a} depend on the type of matrix operation. Matrix multiplication admits a solution as long as the dimensions of the input matrices are compatible. 
However, if matrices represent a system of linear equations that cannot be satisfied simultaneously, the task has no solution~\cite{linear_algebra}, thus invalidating \btp[noname]{timeliness}{b} and possibly \btp[noname]{switch}{a}. Because of this ambivalence, both \btp[noname]{timeliness}{b} and \btp[noname]{switch}{a} depend on the specific class of tasks.
A similar consideration applies to \btp{timeliness}{c}. Assuming \btp[noname]{timeliness}{b} is satisfied, \btp[noname]{timeliness}{c} can be obtained in many matrix-based computations by decomposing the task into sub-problems.

The proposal does not discuss \btp{variability}{}. While the shuffling service may help, it remains unclear in practice whether \btp[noname]{variability}{} is satisfied. The paper suggests that \btp{adj}{} may be attained by batching submitted tasks together. 
On the other hand, difficulty reduction \btp{adj2}{} can be achieved by assigning sub-problems to miners, as argued for \btp[noname]{timeliness}{c}. 
In the case of a new block reception, however, the change of context invalidates the current task, and the average time to find a solution to the new task is higher. This is detrimental to \btp{switch}{c}, and thus miners may not be incentivized to accept incoming blocks.
The proposed system includes the role of auditors, which validate proposals. \btp{sound}{} is therefore satisfied as invalid proposals are likely to be rejected in proportion to the number of verifications performed by the auditors.

The system includes a probabilistic multi-step verification algorithm involving the shuffling service and multiple auditors, with each auditor concurrently verifying a portion of the solution. While dividing the verification among multiple auditors speeds up the process, efficiency in practice remains an open question.
On the other hand, the verification algorithm satisfies \btp{compl}{}, as valid solutions are never rejected.

\btp{fairness}{a} is not considered in the work. In practice, the impact of shuffling on fairness remains unclear. We observe that common solution algorithms for matrix operations are deterministic, thus favoring the centralization of the computation. 
A drawback of matrix-based tasks is the large input size that is necessary for computation hardness, affecting both storage and bandwidth requirements.

\subsection{Solution search while mining}

In this group, proposals retain the use of cryptopuzzles while expanding the block proposal process to incorporate the exploration of solutions for useful tasks. For example, candidate cryptopuzzle nonces can be derived from valid solutions to the useful tasks or a solution to the cryptopuzzle is used as a seed for solving the useful task.
In both cases, miners are motivated to explore a vast solution space of the useful task.
Proof of Search~\cite{pos}, Ofelimos~\cite{ofelimos}, Axechain~\cite{axechain}, Proof-of-Evolution~\cite{bizzarro}, mPoW~\cite{mpow}, as well as the system of~\cite{lihu} belong to this category.

\textbf{Proof of Search} (PoS) is a system where solvers explore the solution space of useful optimization problems while solving cryptopuzzles to propose blocks. 
First, solvers prepare a block template containing transactions, etc. Then, solvers start testing different nonces so that the output of the hash function is below an established difficulty threshold, similar to Bitcoin cryptopuzzles. Unlike Bitcoin, however, PoS nonces are not random sequences of bits; instead, they must include a solution (not necessarily good wrt. the optimization metric) to an optimization problem. 
Additionally, nonces include a context-sensitive string derived from the solution and from the block context. The work proposes a complicated scheme to this end, which consists of computing an optimization metric for the solution and skewing this metric based on the block context. However, we believe that the same goal can be achieved in a simpler way and with stronger guarantees by applying a \emph{contextualizer} function. All solvers and validators must agree upon the contextualizer function; it can, e.g., be provided along with the task. The simplest implementation of the contextualizer function would be a secure hash.

Due to the hardness of cryptopuzzles, many different solutions are tested while proposing a block, increasing the likelihood of finding a good solution with respect to the optimization metric.
To prevent a solution from being stolen by other nodes, PoS uses a mechanism akin to a commitment scheme where every node first registers on the chain the best value found. The way it is implemented is not specified in the paper, however. After a few blocks, each node checks whether its solution is the best, and if so, it discloses the solution. Thus, suppliers receive the best solution found by the whole network.

Unfortunately, the conflict resolution rule is not presented.
At the same time, the work provides an interesting discussion of the incentive model. In particular,
the system assigns a reward for proposing a block and a separate reward, paid by the supplier, for the best solution disclosed. The work explains why the distribution of rewards disincentivizes suppliers and solvers to collude.
Suppliers submit tasks as transactions that are included in the blockchain, and any node can be a supplier. However, how miners select which tasks to solve is not discussed.
PoS does not define usefulness, yet we classify it as strong usefulness since the proposed solution complies with Definition~\ref{def:strong_u}. 
Miners dedicate a portion of their computational efforts to finding a useful task solution.
The proportion of computation spent on useful tasks as opposed to cryptopuzzles highly depends on the solution search algorithm (which is solver is free to choose) as well as on the relative difficulty of the optimization problem, making it difficult to generalize. 
Nevertheless, we note that miners may prefer cheaper algorithms, such as random permutations of TSP cities, over better quality producing but slower and more expensive algorithms.
This dichotomy may lead to the generation of lower-quality solutions overall.

Most BTPs are trivially satisfied by the use of cryptopuzzles; however, we have identified a few that are affected by the introduced nonce scheme.
Among the satisfied properties, \btp{rate}{a} and \btp{variability}{} follow naturally.
\btp{rate}{b} leverages the contextualizer function and is supported like in Bitcoin. 
Similarly, \btp{rate}{c}[fairness][b] follows from Bitcoin.

The verification leaves no probability of invalid proposals being accepted, thus supporting \btp{sound}{}.
Besides, the proposed verification algorithm is both efficient and complete, satisfying both \btp{verif}{b} and \btp{compl}{}.
By assuming that finding a solution to the useful task is cheap compared to cryptopuzzles, \btp{fairness}{a} derives from cryptopuzzles. In practice, such cost depends on several factors, such as the choice of useful tasks and the available solution algorithms; it remains unclear whether PoS satisfies \btp[noname]{fairness}{a}.

An unexpressed assumption of PoS is that the solution space of the useful task instance is large enough to include a valid cryptopuzzle solution.
Imagine the trivial example of a TSP of three cities: there exist only six possible arrangements, hence no more than six nonces to test.
We observe, therefore, that an instance with a small solution space may have a very low probability to yield any valid block. This potential issue is exacerbated by the fact that suppliers may be malicious and supply small instances on purpose.
This is a challenge for \btp{timeliness}{b} and \btp{timeliness}{c}, with the consequences described in Section~\ref{sec:security_properties}. 
The main difference from cryptopuzzles in Bitcoin is that the size of the nonce space in Bitcoin is fixed and controlled by the system while the solution space of the useful task instance in PoS is outside the system control.
The size of the solution space affects \btp{adj}{} and \btp{adj2}{} asymmetrically. \btp[noname]{adj}{} is guaranteed regardless, since it is always possible to arbitrarily increase the difficulty of the cryptopuzzle. However, \btp[noname]{adj2}{} is not satisfied. In particular, this is true for cases where there is no solution because it is not possible to reduce the difficulty of unsolvable instances. Sudden variations in the size of the solution space across different instances may also have an unpredictable effect on \btp{switch}{a} and \btp{switch}{c}.
Besides, the committing scheme to disclose the solutions is likely to face serious scalability issues due to the high number of tasks and nodes.

\textbf{Ofelimos} is a blockchain protocol whose consensus is built around a stochastic local search algorithm called Doubly Parallel Local Search (DPLS). Stochastic local search is a generic algorithmic paradigm for solving optimization problems, and DPLS is designed to be widely applicable to many classes of optimization tasks. 
The core of the mining algorithm consists of executing the DPLS, which constitutes the useful part of the PoW. However, to defend against pre-computation attacks and cherry-picking instances of low complexity, miners must first solve a cryptopuzzle of moderate difficulty $T_1$. The resulting nonce value is used as a seed for the DPLS algorithm, thereby randomizing the computation. 
Miners feed the DPLS output into one single post-hashing round that decides, against an adjustable threshold $T_2$, whether the block is eligible for publication. Note that the miner only learns whether a PoW attempt is successful after executing DPLS, i.e., the computation cannot be cut short to speed up block creation.
A miner repeats the computation sequence of solving a cryptopuzzle, performing useful work, and calculating one post-hash, until the post-hash of a sequence lies below $T_2$, allowing for the block to be published. 
The authors estimate that in the common case, less than half of the total computation is spent on the useful task, while the rest is dedicated to cryptopuzzles.
However, they note that it is possible, in restricted cases related to DPLS hardness, to increase this fraction arbitrarily close to one.

A block contains two points explored using the DPLS algorithm: the one that leads to a sufficiently small post-hash and enables block creation, and the best one in terms of the optimization metric. The latter point is used to progress the DPLS algorithm. 

Ofelimos incorporates two additional optimizations: (a) the cost of block verification is reduced by having miners append a Succinct Non-interactive ARgument (SNARG) proving the correctness of both exploration points contributing to the block, and (b) a 2-for-1 scheme~\cite{garay,fruitchain} is employed to publish local search results faster, without waiting until the block is proposed.
The useful tasks to be solved are selected locally based on public information posted on the blockchain; however, technical details are not presented in the work.
The supply model is not described either. Nevertheless, we classify usefulness as strong since task instances are not modified.

Many BTPs stem from the cryptopuzzle-based pipeline. 
\btp{rate}{a} and \btp{timeliness}{c} follow from the individual hardness of both cryptopuzzles and DPLS, though the work does not provide an exact analysis of the combined hardness.
Tasks are solvable, satisfying \btp{timeliness}{b}, because miners in Ofelimos can always start the cycle of cryptopuzzle, DPLS, and cryptopuzzle again so that eventually, a solution would be found.
Because of the solvability, \btp{switch}{a} holds as well.
\btp{switch}{c} is satisfied when the same values of $T_2$ are consistently used, resulting in the same average time to find a solution.
Increasing or decreasing the value of $T_2$ may help with supporting \btp{adj}{} and \btp{adj2}{}, respectively. However, the exact effect of a given change in $T_2$ on the difficulty of the entire BT, which consists in the three-step pipeline, is difficult to calculate or predict.

\btp{variability}{} follows due to the stochastic nature of DPLS and the well-studied variability of cryptopuzzles. 
Finally, the pre-hashing step ensures the context-dependency of \btp{rate}{b}.

\btp{rate}{c}[fairness][b] is guaranteed for the hashing part, and it is likely satisfied for DPLS because of the pre-hashing step, which prevents miners from cherry-picking useful computations.

The protocol requires the computation of a SNARG to secure DPLS solutions and verifies the result in polynomial time. 
To this end, Ofelimos assumes a bounded number of computation steps to produce and verify the SNARG. No discussion is provided on how realistic such an assumption is given the available SNARG implementations, and thus the impact on \btp{verif}{b} remains unclear. 
\btp{sound}{} and \btp{compl}{} are not discussed, but we conclude they are both satisfied due to the properties of SNARGs~\cite{groth} and cryptopuzzles.
Finally, the authors claim that any set of solvers will produce a subset of blocks roughly proportional to their computing power, resembling \btp{fairness}{a}. However, details and proofs are omitted from the work.


\section{Research gaps and outlook to the future}
\label{sec:outlook}

In the following, we point out selected research directions and gaps.

    \textbf{Formalization:}
    This work provides extensive, albeit somewhat informal, reasoning for the importance of cryptopuzzle properties. An extension to the current work may formalize the BPPs as this paper did with BTPs, as well as include formal proofs of sufficiency and necessity.
    Furthermore, the analysis summarized in Tables~\ref{table:works_security} and~\ref{table:works_liveness} shows that none of the proposed systems satisfies all the identified properties, corroborating the perceived difficulty of the Proof-of-Useful-Work problem. 
    We conjecture that the classes of tasks, such as cryptopuzzles, that satisfy all of the properties are those whose most efficient solution algorithm is a brute-force search.
    In other words, the solution search algorithm can be modeled as a perfect random number generator. 
    However, this conjecture remains unproven for the time being.

    \textbf{Characterization of property support:}
    It is a challenge to draw definitive conclusions on the BTP support given a specific class of tasks. In fact, unlike cryptopuzzles, such analysis must account for the heterogeneity of solution algorithms and problem instances. 
    While NP-complete problems are provably hard, there exists no equivalence class in terms of variability (\btp[noname]{variability}{}) or fairness (\btp[noname]{fairness}{}), to name a few.
    The BTPs we identified in this work can be evaluated experimentally only by introducing assumptions and restrictions on the solution algorithms and instance sets.
    In addition, those types of evaluations are not common in the current research literature, which rather focuses on other types of metrics.
    For example, heuristics for the TSP show extensive results on the solution quality~\cite{tsp_quality1, tsp_quality2} as well as the runtime~\cite{tsp_runtime1} without, generally, analyzing variability. 
    ML literature elaborates extensively on the model accuracy~\cite{aimodernapproach, bishop} and convergence~\cite{bert} while providing little insight on, for example, how the performance varies with the allocated computing power.
    Therefore, we believe there is room for significant new research to investigate less explored metrics and characterize the property support of state-of-the-art algorithms. 
    
    \textbf{Reliable estimation and difficulty adjustment mechanisms:}
    While NP-complete problems are provably hard, not all instances are hard to solve in practice.
    One of the features of cryptopuzzles essential in the context of PoW is that, given a fixed difficulty, all cryptopuzzle instances require, on average, the same solution time. Consequently, solution time is reliably adjusted by changing difficulty. 
    The concept of difficulty in useful tasks, however, is not sufficiently defined. What parameter should be adjusted to increase the solution time e.g. twofold? How does this affect all instances?
    Systems such as~\cite{exercise} invoke the support of a practical mechanism to estimate the solution time. 
    We note that the soundness and availability of reliable and practical mechanisms to estimate the average solution time would greatly strengthen proof of useful work proposals.
    The benefit of having reliable estimates is multifold: such estimates can drive task selection policies to improve, for example, \btp[noname]{fairness}{}. 
    They can also help set the best validity threshold, which is a key issue for many important classes of tasks, as described in Section~\ref{sec:opt_and_decision}. 
    
    \textbf{Models:}
    We observe insufficient exploration of (1) comprehensive supplier models and task selection policies and (2) economic models and incentive mechanisms.
    As extensively discussed, classes of tasks vary broadly wrt. property coverage. 
    However, the analysis of such coverage requires clear models and policies. 
    Nodes may be honest, malicious, or rational.
    Task supply can be external or internal; task selection coordinated or uncoordinated.
    None of the proposals surveyed in Sections~\ref{sec:existing_approaches} and~\ref{sec:related_work} comprehensively cover the possible supplier models and task selection policies, or discuss the trade-offs between various possibilities. 
    While this work has taken the first step toward (1), there is still much work left to be done.
    We believe that such an analysis is necessary to complement the landscape of design choices.


\section{Related work}
\label{sec:related_work}

The term \textit{proof-of-work} preceded blockchains and was first formalized by M. Jakobsson and A. Juels in 1999~\cite{pudding}.
In the PoW scheme, a prover demonstrates to a verifier that a certain amount of computational effort has been expended in a specified time interval. Among the proposed PoW schemes~\cite{pow_wiki, pow}, hashcash~\cite{hashcash} is widely believed to have inspired Bitcoin's cryptopuzzles. 
Since then, repurposing Bitcoin's cryptopuzzles for useful tasks has been an open problem.
However, comparatively little effort has been directed into clearly defining cryptopuzzle properties and analyzing the challenges involved in its replacement.

High-level property analysis has been conducted in~\cite{fair-pouw, dotan, salhab, vanity}, while a more systematic, but still partial, analysis is presented in the book by Narayanan et al.~\cite{bitcoinbook}.
At the time of writing, the most complete list of properties has been proposed by Ball et al.~\cite{ball1}, which also provides a task interface that inspired the interface defined in Section~\ref{sec:base_interface}. Nonetheless, a few important properties are not considered in~\cite{ball1}, most notably decentralization and fairness, as well as switchability. 
In summary, to the present day, there exists no commonly accepted list of properties that blockchain classes of tasks need to satisfy.
To the best of our knowledge, the present work includes the most complete definitions of properties, and it is the first to critically discuss their interplay in a systematic and extensive way.

The property framework developed in the current work is used to critically assess proposed systems in the literature.
Indeed, there are many proposed systems and candidates for cryptopuzzle replacement.
Useful tasks include, but are not limited to, prime number operations~\cite{primecoin, gapcoin}, OV and linear algebra calculations~\cite{ball1, exercise}, Monte Carlo algorithms~\cite{hepchain, nooshare}, storage-intensive searches~\cite{permacoin}, DNA sequence alignment~\cite{coinami}, and generic tasks~\cite{axechain, rem}.
The most popular class of tasks in the proposed works are optimization tasks~\cite{pos, ofelimos, crowdsourcing, coop-hybrid, todorovic} sometimes restricted to the special case of ML, DL and RL~\cite{dlchain, coinai, merlina, felipebravo, pole, podl, biomedical, lihu, dlbc, mittal, q-learning, evolved}, Federated Learning~\cite{proof-of-federated-learning, fedcoin}, and
Genetic Algorithms~\cite{bizzarro, mpow}.
Due to the required hardness, some systems employ NP-complete problems~\cite{hybrid, ricottone1, transportation, chrisimos}, among which TSP is the most frequently chosen class~\cite{generals, evo-pow, adapting}.

Finally, some works propose specific novel techniques rather than a complete system: a difficulty adjustment mechanism~\cite{ricottone2}, a reward system~\cite{prestige}, and specific supply models~\cite{coinami, mittal, fedcoin}. A large body of research focuses on replacing PoW, e.g., using a trusted execution environment~\cite{rem} or storage-based incentives~\cite{permacoin, felipebravo}. Despite these important efforts, PoW remains a highly popular and relevant mechanism.

\section{Conclusion}
\label{sec:conclusion}
The possibility of using the extensive computing power supplied by blockchain miners is tantalizing and motivates the development of our property framework.
This paper identifies and defines several important blockchain properties. We discuss the relevance of each property and the relation between them, highlighting trade-offs and challenges when replacing cryptopuzzles with a different class of tasks. 
The property framework has been used to assess the pitfalls and challenges of several blockchain designs that replace cryptopuzzles with useful work. 
In several cases, the proposed designs address a subset of our properties. 
Our analysis reveals that usefulness is in direct conflict with many important properties. Moreover, no alternative task that can satisfy all properties and completely replace cryptopuzzles has been proposed.
With our framework, we hope to provide standard requirements for useful blockchain developers and researchers and to spark additional research efforts in this direction.

\begin{acks}
We are thankful to many researchers who proofread an earlier version of this survey and provided insightful comments. A complete list of acknowledgments will be provided in subsequent versions.
\end{acks}

\bibliographystyle{acm/ACM-Reference-Format}
\bibliography{references}

\appendix

\section{Matrix size for given computing power and time}
\label{appendix:derivation_matrix_size}

The goal is (a) to estimate the instance size required to propose a k-OV solution every 10 minutes on average, like in Bitcoin, and (b) to demonstrate that the required size is impractical.
While there is no clear unique way to demonstrate (b), every time we select an underlying assumption, we choose the assumption resulting in a smaller instance size i.e. which is unfavorable to (b). This allows us to calculate a lower bound on the actual instance size, and show its impracticality. Under realistic assumptions, that size will likely be even more significant.
 
To demonstrate (b), we start by considering a mining pool and estimate the following: (1) the computing power of the mining pool measured in FLOPS instead of hashes/s, and (2) the expected time for the pool to solve a k-OV task. 
To address step (1), we retrieved updated performance data about CPUs and ASICs.
The latest generation of Intel processors executes roughly $800\cdot10^{9}$ FLOPS~\cite{intelstat}, while the latest AntMiner, a popular mining ASIC, has a hashrate of $190\cdot10^{12}$ hashes/s~\cite{antstat}. 
Directly converting hashes/s into FLOPS leads to an improvement of three orders of magnitude per computing unit. Instead, we select the conservative case of an equal number of computing units, an assumption unfavorable to point (b).

At the time of writing, Bitcoin's biggest mining pool has 100 Ehashes/s\footnote{https://btc.com/stats/pool}. Considering an AntMiner of the latest generation, we estimate that such a mining pool comprises roughly $5\cdot10^{5}$ ASICs. This corresponds to a mining pool of 400 PFLOPS. 

As step (2), we now estimate the expected proposal time.
In Bitcoin, the whole network proposes a new block every 10 minutes on average. Individual mining pools, on the other hand, propose blocks much less frequently. In this example, we assume that the proposal frequency of the mining pool used in our calculations is the same as the proposal frequency of the whole network.
This simplification is unfavorable to the hypothesis in (b).


Finally, by using the upper bound $(ct)^{1/3}$ on the size of matrices,
we estimate that a mining pool of 400 PFLOPS requires matrices of dimension $n \approx 6.2\cdot10^{6}$ to solve a k-OV instance in 600 seconds.

\end{document}